\documentclass[aps,amsfonts,prx,twocolumn,showpacs, superscriptaddress, nofootinbib]{revtex4-2}

\usepackage{graphicx} % Required for inserting images
\usepackage{outlines, enumitem}
\usepackage{xcolor}
\usepackage{hyperref}
\usepackage{physics}
\usepackage{dsfont}
\usepackage{tikz}
\usepackage{quantikz}
\usepackage{amsmath}
\usepackage{amssymb}
\usepackage{soul}
\usetikzlibrary{arrows.meta, positioning, backgrounds, calc, fit}
\setenumerate[1]{label=\arabic*}
\setenumerate[2]{label*=.\arabic*}
\setenumerate[3]{label*=.\arabic*}
\setenumerate[4]{label*=.\arabic*}

\definecolor{tealgreen}{HTML}{00856b}
\definecolor{purple}{HTML}{9403fc}
\definecolor{winered}{HTML}{540B0E}
\definecolor{brownred}{HTML}{9E2A2B}

\definecolor{teal}{HTML}{397f71}   % primary accent
\definecolor{teallt}{HTML}{E3ECE9} % pale green-grey
\definecolor{red}{HTML}{9E2A2B}    % dark red
\definecolor{redlt}{HTML}{F6EDED}  % pale pink
\definecolor{gold}{HTML}{E09F3E}   % ochre
\definecolor{cream}{HTML}{F9F5EA}  % cream
\definecolor{ink}{HTML}{2B2B2B}

\newcommand{\hdr}[1]{{\small\bfseries\color{teal}#1}}
\newcommand{\hdrR}[1]{{\scriptsize\bfseries\color{red}#1}}
\newcommand{\hdrG}[1]{{\scriptsize\bfseries\color{gold!72!black}#1}}

\hypersetup{colorlinks=true, allcolors=tealgreen}
\selectfont
\newif\ifshowcomments
\showcommentstrue

\usepackage{xcolor}

\begin{document}
\pagenumbering{arabic}

\title{Resilience Beyond the Light Cone: Error-Detected Primitives for Practical Dynamic Circuits}
\author{Kevin C. Smith}
\email{kcsmith@ibm.com}
\affiliation{IBM Quantum, IBM Research Cambridge, Cambridge, MA 02142, USA}

\author{Bibek Pokharel}
\affiliation{IBM Quantum, Thomas J. Watson Research Center, Yorktown Heights, NY  10598, USA}

\author{Satvik Maurya}
\affiliation{IBM Quantum, Thomas J. Watson Research Center, Yorktown Heights, NY  10598, USA}
\affiliation{University of Wisconsin-Madison, Department of Computer Sciences, Madison, WI 53706, USA}

\author{Maika Takita}
\affiliation{IBM Quantum, Thomas J. Watson Research Center, Yorktown Heights, NY  10598, USA}

\begin{abstract}
Dynamic circuits, which augment unitary operations with mid-circuit measurements and classical feedforward, can generate long-range entanglement in constant depth, enabling low-depth primitives ranging from nontrivial state preparation to many-qubit entangling gates. Escaping the light-cone constraints of static unitary circuits, however, comes at a cost: these primitives typically require a number of mid-circuit measurements that scales with system size and that, together with feedforward latency, can introduce errors that degrade the long-range entanglement on which they rely. Here, we alleviate this tension by showing that many such primitives, when cast into a common operational framework, admit an error-detection scheme that trades infidelity for a postselection overhead, with no additional ancillas and only constant gate depth. Our framework thus unifies and upgrades a broad class of primitives including fan-out gates, multi-qubit Pauli rotations, the preparation of W and higher-weight Dicke states, and of certain non-normal matrix product states. We also introduce a reduced-depth, error-detected implementation of the Hadamard test, extending the use cases of dynamic circuits to a key algorithmic primitive.
Finally, we establish the practical utility of our scheme through experiments on a superconducting quantum processor. We demonstrate the error-detected preparation of a long-range entangled Bell pair spanning a 100-qubit chain with fidelity $F=0.59\pm0.02$, surpassing the entanglement-certification threshold $F > 0.5$ that the baseline dynamic-circuit implementation fails to reach ($0.39\pm0.01$). Separately, we demonstrate the constant-depth preparation of W states of up to 20 qubits by consuming GHZ states of up to 40 qubits, finding that error detection yields absolute fidelity improvements of $\Delta F\approx0.2$ across the largest sizes studied. Altogether, these results bring low-depth dynamic-circuit primitives within practical reach on present-day hardware.

\end{abstract}

\maketitle

\section{Introduction}

By combining local unitary gates, mid‑circuit measurements, and feedforward operations, dynamic quantum circuits have emerged as a powerful extension to the local unitary circuit model. A key advantage is their ability to spread correlations across arbitrarily many qubits using only a constant‑depth circuit, i.e., one whose depth does not grow with system size; whereas the correlations generated by constant-depth unitary circuits are confined to a finite-width, Lieb-Robinson-like ``light cone''~\cite{Bravyi_LiebRobinsonBounds_2006}, constant-depth dynamic circuits can evade this restriction~\cite{Piroli_QuantumCircuits_2021, Lu_MeasurementShortcut_2022, Verresen_EfficientlyPreparing_2022, Smith_DeterministicConstantDepth_2023}. This capability is especially appealing in the near term, where noise and gate imperfections limit current quantum processors to shallow circuits.

 Motivated by this observation, recent works have leveraged dynamic circuits to develop a variety of constant-depth state preparation and multi-qubit gate protocols that, in a purely unitary framework, would require circuit depth scaling with system size. Examples of the former include certain topological orders~\cite{Verresen_EfficientlyPreparing_2022, Lu_MeasurementShortcut_2022, Tantivasadakarn_HierarchyTopological_2023, Tantivasadakarn_ShortestRoute_2023, Tantivasadakarn_LongRangeEntanglement_2024, Iqbal_TopologicalOrder_2024, Iqbal_NonAbelianTopological_2024}, matrix product and projected entangled pair states~\cite{Smith_DeterministicConstantDepth_2023, Smith_ConstantDepthPreparation_2024, Sahay_ClassifyingOneDimensional_2025, Stephen_PreparingMatrix_2025, Zhang_CharacterizingMatrixProduct_2024}, and other multipartite entangled states such as W and Dicke states~\cite{Buhrman_StatePreparation_2024, Piroli_ApproximatingManyBody_2024, Farrell_DigitalQuantum_2025}. On the gate synthesis front, dynamic circuits enable constant-depth implementations of long-range two-qubit gates~\cite{Gottesman_QuantumTeleportation_1999, Baumer_EfficientLongRange_2024}, multi-qubit Pauli rotations~\cite{Moflic_ConstantDepth_2025, Yang_HarnessingPower_2024}, arbitrary Clifford circuits~\cite{Jozsa_IntroductionMeasurement_2005, Buhrman_StatePreparation_2024}, and other gate primitives ~\cite{Quek_MultivariateTrace_2024, Piroli_ApproximatingManyBody_2024, Baumer_MeasurementbasedLongrange_2025, Foxman_RandomUnitaries_2025, Goldstein-Gelb_COMPASDistributed_2026}. A notable instance is the fan-out gate, which underlies constant-depth constructions of the $N$-qubit Toffoli gate and the quantum Fourier transform~\cite{Hoyer_QuantumFanout_2005, Buhrman_StatePreparation_2024, Gretta_ShorsAlgorithm_2026}, and which plays a central role in recent proposals to use dynamic circuits for applications such as quantum imaginary time evolution~\cite{Lund_ConstantDepthQuantum_2026}, sparse state preparation~\cite{Yeo_ReducingCircuit_2025}, and measurement-driven quantum advantage in shallow circuits~\cite{Cao_MeasurementDrivenQuantum_2026}. 

However, these reductions in circuit depth come with important tradeoffs. First, there is a well-appreciated spacetime tradeoff between unitary and dynamic realizations of the same task: while the latter can reduce depth, it typically requires at least $O(N)$ ancillas\footnote{For certain global primitives, such as $N$-qubit Toffoli gates, Clifford grid circuits, and the quantum Fourier transform, the ancilla overhead can grow even faster, scaling as $O(N\log{N})$, $O(N^2)$, and $O(N^3\log N)$, respectively~\cite{Buhrman_StatePreparation_2024}.}. Less emphasized, however, is a second cost: fully collapsing the light cone requires $O(N)$ mid-circuit measurements ~\cite{Friedman_LocalityError_2023} which -- together with the associated feedforward latency -- can introduce errors that substantially degrade the long-range entangling power of dynamic circuits. On superconducting platforms, for example, readout errors are often a dominant noise source and can exceed two-qubit gate errors. In addition, dynamic circuit protocols are limited in their ability to reduce two-qubit gate count relative to their unitary gate count, as local operations and classical communication (LOCC) cannot create entanglement. Altogether, these features create a tension between the theoretical promise of dynamic circuits and their practical performance on noisy hardware.

In this work, we address this tension by introducing a general, low-overhead error-detection strategy that improves the resilience of common dynamic-circuit primitives to readout and gate errors through postselection. Our framework relies on several key advances. First, we identify \emph{distributed control}, wherein a single logical control is encoded nonlocally into a GHZ state, as a common organizing principle underlying a broad class of recently developed constant-depth dynamic-circuit constructions. While related ideas have long appeared in other settings, including fault-tolerant syndrome extraction~\cite{Shor_FaulttolerantQuantum_1996, DiVincenzo_FaultTolerantError_1996}, quantum fan-out circuits~\cite{Moore_ParallelQuantum_1998, Hoyer_QuantumFanout_2005}, and distributed quantum computing~\cite{Yimsiriwattana_GeneralizedGHZ_2004}, here we recontextualize these ideas in the setting of dynamic circuits. In particular, we introduce two key primitives -- \textsc{distribute} and \textsc{collapse} -- that enable the constant-depth distribution and reduction of logical control using measurement and feedforward. We then show that these primitives unify a broad class of recently developed protocols, including constant-depth implementations of long-range and many-qubit gates~\cite{Baumer_MeasurementbasedLongrange_2025}, as well as the preparation of certain long-range entangled states such as W and Dicke states~\cite{Piroli_ApproximatingManyBody_2024}, structured towers of excited states~\cite{Guo_TowerStructured_2026}, and non-normal matrix product states~\cite{Smith_ConstantDepthPreparation_2024}.

Second, we show how to augment \textsc{collapse} with two low-overhead error detection strategies. In the first, we rely on a set of measurements we refer to as \emph{explicit} checks, which enable detection of both fusion-measurement and gate errors during the preparation and use of the GHZ state. Our second strategy, which leverages \emph{implicit} checks, is more powerful: it detects all errors captured by explicit checks in addition to measurement errors during \textsc{collapse}. Consequently, as we will show, it effectively trades readout-induced infidelity for classical postselection overhead.

Third, as a new application of these ingredients, we show how they can be combined to realize a low-depth, error-detected implementation of a key primitive in quantum algorithms: the Hadamard test. In particular, for a Hadamard test of an $N$-qubit unitary $U$ that decomposes into $k$ layers of single- and two-qubit gates, a standard single-ancilla implementation requires depth $O(Nk)$ in the worst case. In contrast, our approach requires a dynamic circuit of depth $O(k)$ with only a single round of mid-circuit measurement and feedforward, while incorporating built-in error detection. Applications of the Hadamard test are far-ranging~\cite{Faehrmann_ShadowHadamard_2025} and include, for example, phase estimation and related approaches to spectral estimation~\cite{Kitaev_QuantumMeasurements_1995, Somma_QuantumEigenvalue_2019, Parrish_QuantumFilter_2019, Shen_EstimatingEigenenergies_2025}, dynamical correlation functions and linear response~\cite{Somma_SimulatingPhysical_2002, Roggero_DynamicLinear_2019, Wang_QubitEfficientRandomized_2024, Cruz_QuantumSimulation_2025}, entanglement spectroscopy~\cite{Johri_EntanglementSpectroscopy_2017}, and trace and fidelity estimation~\cite{Knill_PowerOne_1998, Buhrman_QuantumFingerprinting_2001}. Our reduced-depth, error-detected implementation therefore brings this wide array of applications within practical reach of near-term devices.

Before proceeding, we make a few remarks regarding prior work. In Ref.~\cite{Liao_AchievingComputational_2025}, the authors demonstrated a teleportation-based long-range \textsc{cnot} gate augmented with a unitary ``entangle-disentangle'' error-detection protocol. The main idea is to combine efficient compilation based on GHZ state injection~\cite{Yang_HarnessingPower_2024} with the observation that, after use, the GHZ state can be reduced in size by unitarily disentangling qubits. This then introduces an error-detection opportunity, since one can check whether the disentangled qubits return to $\ket{0}$. However, this protocol remains limited in two important respects. First, it involves the linear-depth unitary preparation (and unpreparation) of a GHZ state, thereby sacrificing the constant-depth advantage of a measurement-based approach for the ability to detect bit-flip gate errors\footnote{We note that Ref.~\cite{Liao_AchievingComputational_2025} discusses potential routes for depth reduction through hybridization of unitary and measurement-based GHZ preparation. However, they suggest that extending error detection to the full constant-depth measurement-based protocol would require weight-three checks; we show that this is not necessary.}. Ideally, one would have a scheme that retains constant depth while detecting not only gate errors, but also the readout errors that limit the practical utility of dynamic circuits. Second, its construction is developed for a specific primitive; while extensions to fan-out gates and the teleportation of multi-qubit gates are briefly discussed, a broader framework is lacking. These observations motivate the general framework developed here, which introduces error-detection techniques tailored to constant-depth, measurement-based dynamic-circuit primitives and extends their reach to a broad class of gates, state preparations, and key algorithmic primitives such as the Hadamard test. In doing so, our framework also recovers the protocol of Ref.~\cite{Liao_AchievingComputational_2025} as a special limit. 

The remainder of this paper is structured as follows. In Sec.~\ref{sec:error-detection}, we introduce the concept of distributed control, explain our error detection framework at a high level, and provide evidence through noisy simulation that it significantly improves the utility of dynamic circuits that rely on noisy mid-circuit measurements. In Sec.~\ref{sec:primitives}, we show that our approach provides a unified framework for realizing error-detected, constant-depth primitives, and we provide specific examples of both error-detected multi-qubit gates (Sec.~\ref{ssec:gates}) and state preparations (Sec.~\ref{ssec:states}). In Sec.~\ref{sec:experiment}, we present experimental results that demonstrate our framework improves performance of dynamic circuits on a noisy quantum processor. Specific experiments include the implementation of a long-range \textsc{cnot} gate (Sec.~\ref{ssec:lrcx}) and the constant-depth preparation of the W state (Sec.~\ref{ssec:wstate}). We then conclude in Sec.~\ref{sec:conclusion}.

\section{Error detection for dynamic circuits}
\label{sec:error-detection}

\begin{figure*}
    \centering
    \includegraphics[width=1\linewidth]{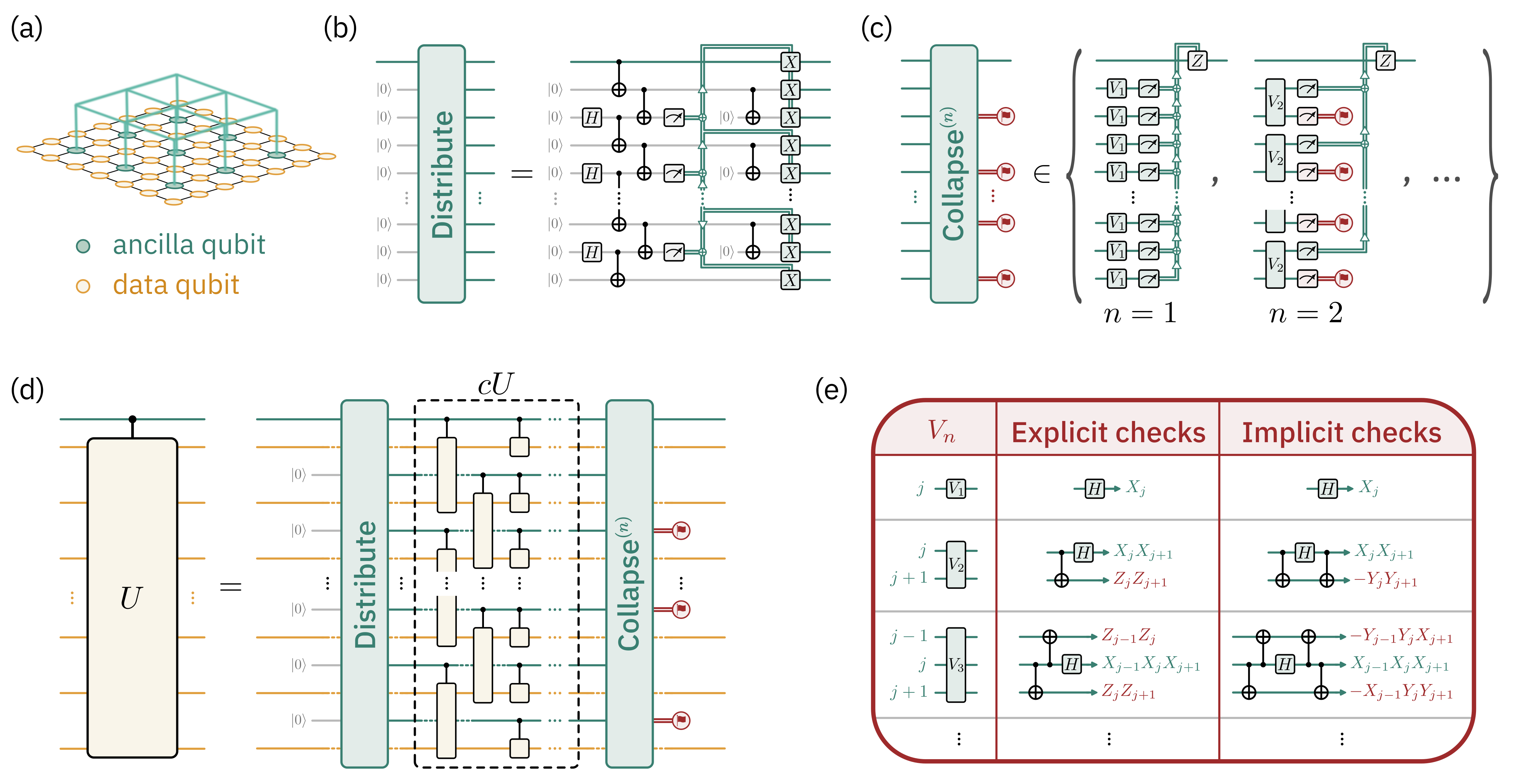}
    \caption{The error-detected \textsc{distribute}--\textsc{collapse} framework. (a) Ancilla qubits are prepared in a GHZ state and distributed throughout the device, forming a \emph{distributed control}. (b) The \textsc{distribute} primitive, which carries out the constant-depth preparation of the distributed control using dynamic circuits. (c) The \textsc{collapse}$^{(n)}$ primitive, which carries out the constant-depth reduction of the distributed control onto a single ancilla. By leveraging the underlying repetition-code-like structure of the distributed control, the baseline implementation ($n=1$) can be upgraded to provide error detection ($n \geq 2$) at the cost of a layer of local check gadgets $V_n$; larger $n$ improves code distance at the expense of higher gate overhead. (d) The error-detected implementation of a controlled global unitary ($cU$) using the \textsc{distribute}--\textsc{collapse} framework. For $U$ of depth $O(k)$, it implements the controlled variant in the same $O(k)$ depth with built-in error detection by localizing and parallelizing the controlled operations. (e) Two types of checks. Explicit checks directly measure the stabilizers $Z_j Z_{j+1}$, flagging fusion measurement errors in \textsc{distribute} and ancilla bit-flip errors from gate and idle noise throughout the circuit. Implicit checks trade additional two-qubit gates for the ability to also detect measurement errors during \textsc{collapse}$^{(n)}$.}
    \label{fig:overview}
\end{figure*}

We begin by considering the concrete task of implementing a global operation $U$ controlled on a single qubit:
\begin{equation}
\includegraphics[width = 0.85\linewidth]{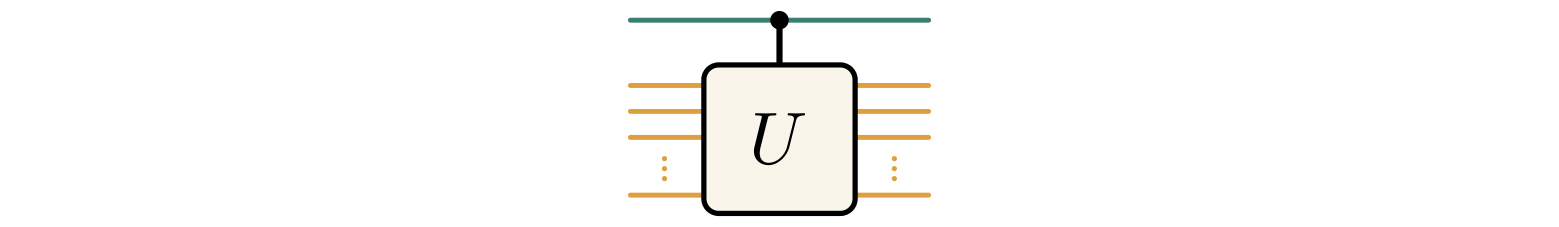}
\label{eq:cU}
\end{equation}
Implementing this operation with local unitary gates requires a circuit depth that scales with system size. In particular, if $U$ can be implemented with circuit depth $k$ then its controlled variant generically requires depth $O(Nk)$ in the worst case as the control qubit can only interact with one target qubit at a time, forcing an effective serialization of each layer. In architectures with connectivity constraints, this overhead can be further compounded by a system-size-dependent \textsc{SWAP} overhead, as the control qubit must be routed to interact with each target.

With additional access to $O(N)$ ancillas, a standard technique for reducing this overhead is to distribute the control across an ancillary GHZ state, thereby enabling the controlled operations to be localized and parallelized across the device~\cite{Moore_ParallelQuantum_1998, Shor_FaulttolerantQuantum_1996, Hoyer_QuantumFanout_2005}. This strategy eliminates the depth overhead due to serialization, shifting the overhead instead to the cost of GHZ preparation and unpreparation. With local unitary circuits alone, this incurs a minimum circuit depth that is at best logarithmic and at worst linear in system size with all-to-all and linear connectivity, respectively. However, it is well-known that GHZ states can be prepared in constant-depth with mid-circuit measurements and feedforward~\cite{Raussendorf_OneWayQuantum_2001, Piroli_QuantumCircuits_2021}. Thus, by leveraging dynamic circuits that support these additional capabilities, the system-size depth dependence can be entirely eliminated. This, in turn, enables the realization of controlled-$U$ in $O(k)$ depth, and forms the backbone of our work. 

We now introduce a pair of primitives -- \textsc{distribute} and \textsc{collapse}$^{(n)}$ -- that make this scheme, illustrated in Fig.~\ref{fig:overview}(d), precise. As we will show, the structure of \textsc{collapse}$^{(n)}$ naturally gives rise to a lightweight yet powerful error-detection scheme. 

\subsection{\textsc{DISTRIBUTE} and ${\textrm{COLLAPSE}^{(n)}}$}
\label{ssec:distribute_and_collapse}

We consider an initial state of $N$+1 qubits written in the factored form,
\begin{equation}
\ket{\Psi} = \alpha \ket{0}_c \ket{\psi_0} + \beta \ket{1}_c\ket{\psi_1},
\end{equation}
where we have separated the state of the control qubit $|j\rangle_c$ from that of the $N$ target qubits $|\psi_j\rangle$.

Our first primitive, \textsc{distribute}, encodes the logical state of the control qubit into a GHZ-like state across $O(N)$ ancilla qubits that extend across the device:
\begin{equation}
    \ket{\Psi} \xrightarrow[]{\textsc{distribute}}|\tilde{\Psi}\rangle.
    \label{eq:distribute}
\end{equation}
Here, the tilde denotes the \emph{distributed} state,
\begin{equation}
|\tilde{\Psi}\rangle = \alpha \ket{\tilde{0}}_c \ket{\psi_0} + \beta \ket{\tilde{1}}_c\ket{\psi_1},
\end{equation}
with $\ket{\tilde{0}} = \ket{000\ldots 0}$ and $\ket{\tilde{1}} = \ket{111\ldots 1}$. We refer to this GHZ-like state interchangeably as the \emph{distributed control} or \emph{distributed ancilla}, as it allows the controlled operations to be localized to individual qubits and executed in parallel. Our specific construction for $\textsc{distribute}$ is shown in  Fig.~\ref{fig:overview}(b), involving a single round of $O(N)$ mid-circuit measurements and feedforward. 

We note that the map in Eq.~\eqref{eq:distribute} is commonly referred to as fan-out. However, our implementation of $\textsc{distribute}$ in Fig.~\ref{fig:overview}(b) is not a quantum fan-out gate, but is instead a \textsc{cnot} ladder.  While both carry out the desired map on the given input state, the constant-depth implementation of the latter requires fewer two-qubit gates and simpler feedforward operations~\cite{Baumer_MeasurementbasedLongrange_2025}. For this reason, we adopt the distinct terminology \textsc{distribute} to differentiate from standard fan-out. We refer to the mid-circuit measurements underlying \textsc{distribute} as \emph{fusion measurements}, in correspondence to the underlying strategy for GHZ preparation: first prepare many small GHZ states in parallel, fuse them together with mid-circuit measurements, and make corrections with feedforward. See Ref. ~\cite{Smith_DeterministicConstantDepth_2023} for details.

Our second primitive, \textsc{collapse$^{(n)}$}, carries out the inverse map. For instance, acting on the distributed state $|\tilde{\Psi}\rangle$, it recovers the state $\ket{\Psi}$:
\begin{equation}
|\tilde{\Psi}\rangle \xrightarrow[]{\textsc{collapse$^{(n)}$}} \ket{\Psi}.
\end{equation}
The superscript $n$ parametrizes the error-detection capability of the primitive: larger $n$ enables stronger error detection at the cost of additional two-qubit gates. We now explain this structure, beginning with the baseline case $n=1$.

\subsubsection{Without error detection: $n = 1$}

The simplest implementation, \textsc{collapse$^{(1)}$}, is shown as the leftmost instance in Fig.~\ref{fig:overview}(c). The central idea is to measure all but one qubit of the distributed ancilla in the $X$ basis, thereby projecting the GHZ state onto a single remaining qubit, up to a measurement-dependent phase that is removed with feedforward. In particular, let $x_{j}$ denote the outcome of measuring $X_j$ and let $x_{N}\ldots x_{3}x_{2}$ represent the bitstring associated with measurements of $\{X_j\}_{j=2}^{N}$ (the choice of leaving qubit $j=1$ unmeasured is arbitrary; any qubit of the distributed ancilla can be selected). Performing these measurements on $|\tilde{\Psi}\rangle$, we obtain
\begin{equation}
\langle x_{N}\ldots x_{3}x_{2}|\tilde{\Psi}\rangle = \alpha|0\rangle_c|\psi_0\rangle + (-1)^{\Pi(x_{N}\ldots x_{3}x_{2})}\beta|1\rangle_c|\psi_1\rangle,
\end{equation}
where $\Pi(x_{N}\ldots x_{3}x_{2}) = \bigoplus_{j=2}^N x_j$ is the parity of the bitstring. Applying $Z^{\Pi(x_{N}\ldots x_{3}x_{2})}$ to the remaining ancilla then recovers $\ket{\Psi}$.

We note that this baseline variant of our primitives, \textsc{distribute} and \textsc{collapse}$^{(1)}$, can be understood as constant-depth, dynamic-circuit based implementations of the cat-entangler and cat-disentangler of Ref.~\cite{Yimsiriwattana_GeneralizedGHZ_2004}, there developed for distributed quantum computing. Our central contribution is the error-detected generalization \textsc{collapse}$^{(n)}$, to which we now turn.

\subsubsection{With error detection: $n\geq 2$}

Several insights underlie the error-detected extension to general $n$. First, in the above procedure we need not resolve each individual bit $x_j$. It suffices to determine the parity of the collective measurement outcomes, $\Pi(x_{N}\ldots x_{3}x_{2})$. From an operator perspective, this corresponds to extracting the joint observable,
\begin{equation}
    \bar{X}_{2:N} = X_N\ldots X_3 X_2.
\end{equation}
Accordingly, $\bar{X}_{2:N}$ can be resolved either by measuring each single-qubit Pauli-$X$ operator individually, as in \textsc{collapse$^{(1)}$}, or by partitioning the product into disjoint $n$-qubit patches and measuring $X^{\otimes n}$ on each. This latter approach underlies \textsc{collapse$^{(n)}$}.

Second, complementary to this perspective is the observation that a physical GHZ state on $N$ qubits can be reinterpreted, without additional gates, as a GHZ state on $N/n$ logical qubits, where each logical qubit is encoded into an $n$-qubit repetition code ($1\leq n \leq N$). Under this reinterpretation, partitioning $\bar{X}_{2:N}$ into size-$n$ blocks corresponds to measuring the associated logical operators $\{
\bar{X}_{s(k) + 1:s(k) + n}\}_{k = 1}^{(N-1)/n}$ where $s(k) = (k-1)n + 1$. Importantly, this requires us to extract only $N_r = (N-1)/n$ bits of information, leaving $N - N_r$ qubits unmeasured. These additional degrees of freedom can then be used to verify the GHZ stabilizers (or, equivalently, those of the underlying repetition code patches), $Z_j Z_{j+1}$.

A final insight motivating this construction is that mid-circuit measurement errors during \textsc{distribute} are effectively converted into bit-flip errors on the GHZ state. Such errors manifest as domain walls, violating the local stabilizer constraints $Z_jZ_{j+1} = +1$. By choosing repetition code patches to overlap the potential locations of these domain-walls, such errors can be detected alongside other sources of bit-flip errors.

The particular error-detection strategy employed by \textsc{collapse}$^{(n)}$ is enforced through choice of the pre-measurement gadget $V_n$. As shown in Fig.~\ref{fig:overview}(b), this operator acts independently on each repetition code patch and facilitates the direct measurement of multi-qubit Pauli operators. We now describe two strategies -- \emph{explicit} and \emph{implicit} checks -- that correspond to different choices of $V_n$.

\subsubsection{Explicit checks}

The simplest choice for $V_n$ encodes the logical operator $\bar{X}_{j+1:j+n}$ on one ``root'' qubit, and explicit stabilizer checks ${Z_j Z_{j+1}}$ on the $n-1$ ``check'' qubits (Fig.~\ref{fig:overview}(e), second column). The no-error condition corresponds to measuring all check qubits to be in $|0\rangle$. As such, this strategy can equivalently be viewed as locally disentangling patches of the GHZ state (requiring $n$ two-qubit gates per patch), leaving behind a smaller GHZ on the $N/n$ root qubits. 

We note that, in the limit $n\to N$, explicit checks can be viewed as a generalization of the ``unitary entangle-disentangle'' protocol of Ref.~\cite{Liao_AchievingComputational_2025}. However, in contrast to that work, here we tailor the stabilizer measurements to detect errors in a distributed control prepared via dynamic circuits. In particular, misassignment errors in the fusion measurements underlying \textsc{distribute} result in domain-wall-like errors in the prepared GHZ state (for example, on six qubits a single fusion measurement error can produce states like $[\ket{000111} + \ket{111000}]/\sqrt{2}$). By aligning detection patches so that the $Z_jZ_{j+1}$ stabilizer checks intersect these domain walls, such errors can be detected and effectively eliminated through postselection.

Beyond fusion measurement errors, explicit checks also detect ancilla bit-flip errors induced by noisy gates or idling throughout the circuit, including during \textsc{distribute}, \textsc{collapse}$^{(n)}$, and any intermediate operations on the distributed ancilla. Dephasing errors are not detected but can be suppressed with techniques such as dynamical decoupling \cite{lidar2014review}. A key limitation, however, is that readout errors in \textsc{collapse}$^{(n)}$ go undetected. Such errors can be fatal: a single readout error among the $N/n$ root qubits can flip the inferred parity, inducing a Pauli-$Z$ error on the remaining control qubit after feedforward. For tasks such as the Hadamard test, this manifests as a bit-flip in the final measurement outcome. We next show how implicit checks additionally detect and eliminate these errors to $(n-1)$th order through postselection.

\subsubsection{Implicit checks}

By slightly modifying the above strategy (and adding $n-1$ additional two-qubit gates per patch), we can also detect readout errors during \textsc{collapse}$^{(n)}$, thereby extending the detection capability to encompass mid-circuit measurement errors in both \textsc{distribute} and \textsc{collapse}$^{(n)}$. 

In this strategy, $V_n$ encodes $\bar{X}_{j+1:j+n}$ on the root qubit (as before) and products of $\bar{X}_{j+1:j+n}$ with stabilizers on the check qubits (see Fig.~\ref{fig:overview}(e)). For example, in the simplest case $V_2$, the root qubit encodes $X_j X_{j+1}$ and the single check qubit encodes $-Y_j Y_{j+1}$. Noting that $-Y_j Y_{j+1} = (X_j X_{j+1})(Z_j Z_{j+1})$, measurements of the root and check qubits should agree whenever the stabilizer is preserved, i.e., $Z_j Z_{j+1} = +1$. We refer to this as an \emph{implicit} check, as the stabilizer outcomes are inferred by comparing multiple measurements.

The key advantage of implicit checks is that they combine stabilizer verification with redundant extraction of $X_{j+1:j+n}$. As a result, errors are flagged not only when stabilizers are violated, but also when measurement outcomes are internally inconsistent, enabling detection of readout errors during \textsc{collapse}$^{(n)}$. While this capability requires twice the two-qubit gates compared to explicit checks, this tradeoff is advantageous when mid-circuit measurements are noisier than two-qubit gates. As we will show in Sec.~\ref{sec:experiment}, we find that implicit checks generally outperform explicit checks.

\subsubsection{Special cases: \textsc{distribute}$^*$ and \textsc{collapse}$^{(n)*}$}
\label{sssec:special}

The full sequence of (i) \textsc{distribute}, (ii) controlled-$U$, (iii) \textsc{collapse}$^{(n)}$ generically requires two rounds of mid-circuit measurement and feedforward. Because feedforward latency can be a significant source of overhead on current hardware, reducing the number of such rounds is desirable. As we show in Sec.~\ref{sec:primitives}, many important primitives can be simplified to require just a single round.

For instance, if controlled-$U$ is Clifford, then the Pauli-$X$ 
corrections within \textsc{distribute} can be propagated 
through both the controlled-$U$ and $V_n$ layers (the latter being Clifford by construction) and absorbed into the feedforward stage of \textsc{collapse}$^{(n)}$. A similar simplification holds when $U$ is Hermitian: the corrections can again be deferred, with the caveat that the feedforward now involves an uncontrolled application of $U$ on the target register. Throughout this work, we use \textsc{distribute}$^*$ to denote the variant in which the feedforward step is deferred.

Likewise, the feedforward stage of \textsc{collapse}$^{(n)}$ can be deferred to post-processing whenever the circuit either terminates with \textsc{collapse}$^{(n)}$ or is followed only by Clifford operations; we denote this variant \textsc{collapse}$^{(n)*}$. As we will discuss in Section~\ref{sec:primitives}, the Hadamard test is one example where this simplification applies, reducing the protocol to a single round of feedforward.

\subsection{The impact of error detection}
\label{ssec:simulation}

\begin{figure*}
    \centering
    \includegraphics[width=1\linewidth]{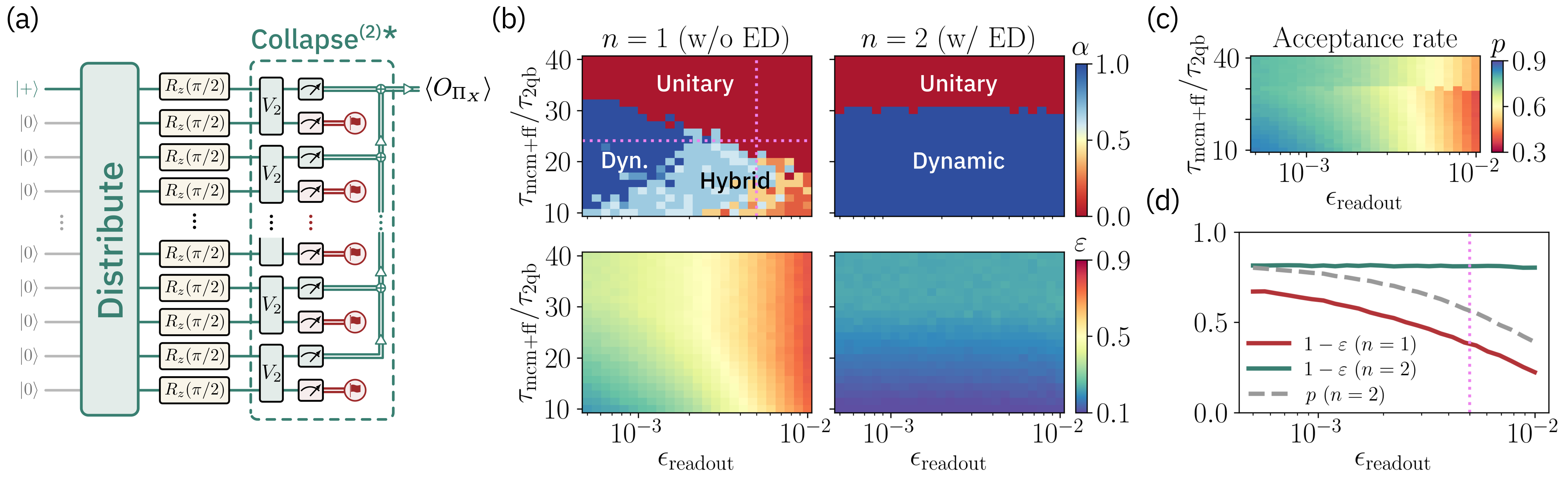}
    \caption{Demonstrating the impact of error detection (ED) through noisy simulation. (a) Circuit diagram for the sensing-like experiment described in Sec.~\ref{ssec:simulation}, involving the preparation (\textsc{distribute}) and consumption (\textsc{collapse}$^{(2)*}$) of a GHZ state to estimate the phase accumulated under an external field through estimation of $\langle O_{\Pi_X} \rangle$. We carry out noisy simulations for $N=52$ qubits using Qiskit's matrix product state backend~\cite{Javadi-Abhari_QuantumComputing_2024} with $T_1 = T_2 = 250\,\mu$s and $\epsilon_{\mathrm{2qb}} = 1.5\times 10^{-3}$, chosen to model \texttt{ibm\_boston}. (b) For each value of mid-circuit measurement and feedforward duration relative to the two-qubit gate duration ($\tau_{\mathrm{mcm+ff}}/\tau_{\mathrm{2qb}}$) and readout error ($\epsilon_{\mathrm{readout}}$), we determine the \textsc{distribute} mid-circuit measurement usage ratio $\alpha$ that minimizes the error $\varepsilon \equiv 1 - \langle O_{\Pi_X}\rangle$, with $\alpha = 1$ and $\alpha=0$ corresponding to constant-depth dynamic and static unitary circuits, respectively, and $0<\alpha<1$ a ``hybrid'' approach that interpolates between the two. Top: The measurement usage ratio $\alpha$, indicating the optimal strategy. Bottom: The corresponding minimized error $\varepsilon$ with (right) and without (left) ED. With ED, we observe $\varepsilon$ to be independent of $\epsilon_{\mathrm{readout}}$, expanding the regime where dynamic circuits are optimal. Dotted pink lines denote the median readout error $\epsilon_{\mathrm{readout}} = 4\times 10^{-3}$ and duration ratio  $\tau_{\mathrm{mcm+ff}}/\tau_{\mathrm{2qb}} = 24.1$ on \texttt{ibm\_boston}, with the latter estimated via circuit scheduling (see App.~\ref{app:noise_model} for details). (c) Acceptance rate $p$ for $n=2$. We observe that $p$ decreases monotonically with increasing $\epsilon_{\mathrm{readout}}$, reflecting the classical sampling cost of rendering $\varepsilon$ independent of $\epsilon_{\mathrm{readout}}$ through postselection. (d) Cross-section of the accuracy $1-\varepsilon$ and acceptance rate $p$ for $\tau_{\mathrm{mcm+ff}}/\tau_{\mathrm{2qb}} = 24.1$ (horizontal dotted pink line, panel (b)).}
    \label{fig:simulation}
\end{figure*}

To illustrate the practical impact of error detection within the 
\textsc{distribute}--\textsc{collapse} framework, we carry out noisy simulations of the circuit shown in Fig.~\ref{fig:simulation}(a) and consisting of the following steps: 
\begin{enumerate}
    \item Prepare an $N$ qubit GHZ state by applying \textsc{distribute} to the input state $\ket{+}$.
    \item Apply $R_z(\pi/2)$ on all qubits.
    \item Estimate the real part of the acquired phase, encoded by $\langle O_{\Pi_X}\rangle = \langle X_1X_2X_3\ldots X_N\rangle$, via \textsc{collapse}$^{(2)*}$.
\end{enumerate}
 Details of the noise model can be found in App.~\ref{app:noise_model}. Taken together, these steps implement a sensing-like experiment in which phase accumulation on a (potentially noisy) GHZ state is used to detect an external field. Several considerations informed the choice of per-qubit phase $\phi = \pi/2$. First, it ensures the overall circuit is Clifford, enabling efficient simulation. Second, the azimuthal rotations $R_z(\pi/2)$ convert individual bit-flip errors into a product of bit- and phase-flip errors which anti-commute with $ O_{\Pi_X}$ and thus maximally degrade the signal, providing an ideal case study for understanding the impact of error detection. Finally, by additionally constraining $N$ to be a multiple of four (here we investigate $N=52$), the total accumulated phase is a multiple of $2\pi$ such that $\langle O_{\Pi_X}\rangle_{\mathrm{ideal}} = +1$.

 To fully capture the impact of error detection, we not only investigate the error in estimating $\langle O_{\Pi_X}\rangle$, but couple this with a more fundamental question: does error detection alter the regime in which dynamic circuits outperform their unitary counterparts? In particular, an alternative to our constant-depth implementation of \textsc{distribute}, which requires $N/2 - 1$ mid-circuit measurements and feedforward, is a linear-depth, fully unitary implementation, requiring zero mid-circuit measurements and no feedforward. More generally, there exists a spectrum of hybrid approaches that involve unitary preparation of $m+1$ GHZ states of size $N/{m+1}$ that are then ``fused'' with $m$ mid-circuit measurements and corrected with feedforward. All strategies in this spectrum are compatible with our error-detected \textsc{collapse}$^{(n)}$ primitive. While dynamic circuits are expected to provide the optimal strategy asymptotically~\cite{Baumer_EfficientLongRange_2024}, for finite-sized systems the additional errors introduced by mid-circuit measurements and the associated feedforward latency can counteract any gains from reduced depth. A natural question, then, is how the optimal strategy varies with the measurement error rate and feedforward latency, and whether error detection reshapes this landscape.

We present our findings in Fig.~\ref{fig:simulation}. In particular, we carry out independent simulations varying three parameters: (i) $\tau_{\mathrm{mcm+ff}}/\tau_{\mathrm{2qb}}$, the combined duration of mid-circuit measurement and feedforward in units of two-qubit gate duration, (ii) $\epsilon_{\mathrm{readout}}$, the error rate for both mid-circuit and terminal measurements, and (iii) $m\in[0, N/2)$, the number of mid-circuit measurements used in \textsc{distribute}, which interpolates between $O(N)$-depth unitary and the $O(1)$-depth dynamic circuit implementations. For each pair $(\epsilon_{\mathrm{readout}}, \tau_{\mathrm{mcm+ff}}/\tau_{\mathrm{2qb}})$ we sweep over all $m$ and identify the value $m_{\mathrm{opt}}$ that minimizes the error $\varepsilon = 1 - \langle O_{\Pi_X}\rangle$. In Fig.~\ref{fig:simulation}(b) we report the measurement usage ratio $\alpha = m_{\mathrm{opt}}/(\frac{N}{2} - 1)$ (top row) and corresponding error $\varepsilon$ (bottom row) with and without error detection. For more information on the system-size dependence of $\alpha$, see App.~\ref{app:noise_model}.

Without error detection, we find that $\varepsilon$ increases monotonically in both $\tau_{\mathrm{mcm+ff}}/\tau_{\mathrm{2qb}}$ and $\epsilon_{\mathrm{readout}}$. Moreover,  $\alpha$ varies substantially across this landscape, with regions of high latency and readout error favoring a unitary strategy ($\alpha = 0$), while strategies involving mid-circuit measurement and feedforward ($\alpha > 0$) become advantageous as $\tau_{\mathrm{mcm+ff}}/\tau_{\mathrm{2qb}}$ and $\epsilon_{\mathrm{readout}}$ decrease. Interestingly, decreasing the former produces a sharp transition from an $O(N)$-depth unitary to an $O(1)$-depth dynamic strategy only when $\epsilon_{\mathrm{readout}} \lesssim 2\times 10^{-3}$; when $\epsilon_{\mathrm{readout}}$ exceeds this threshold, the fidelity becomes readout error limited and a ``hybrid'' approach that trades mid-circuit measurements for unitary depth is favored.

With error detection activated, the optimal strategy shifts decisively toward maximal use of mid-circuit measurements and feedforward below a threshold value of $\tau_{\mathrm{mcm+ff}}/\tau_{\mathrm{2qb}} \approx 30$. Furthermore, the error $\varepsilon$ becomes nearly independent of $\epsilon_{\textrm{readout}}$, with the dependence instead inherited by the acceptance rate which decreases monotonically with increasing $\epsilon_{\textrm{readout}}$. In other words, error detection effectively converts the infidelity resulting from readout error into a classical postselection overhead (see Fig.~\ref{fig:simulation}(c)). This tradeoff is the central principle underlying our error detection strategy, enabling the beneficial use of dynamic circuits in regimes where mid-circuit measurements would otherwise impair their ability to generate long-range entanglement.

Next, we show how our error detection scheme can be used to upgrade dynamic-circuit primitives spanning both multi-qubit gate compilation and state preparation tasks.

\section{Error-detected primitives}
\label{sec:primitives}

\subsection{Multi-qubit gates}
In this Section, we provide a non-exhaustive set of illustrative examples of constant-depth, long-range and multi-qubit gate compilation tasks implementable within our framework. The circuits corresponding to these examples are shown in Fig.~\ref{fig:gates}.
\label{ssec:gates}

\subsubsection{Long-range controlled-U}
Our first example is the implementation of a long-range, controlled single-qubit unitary $U$, shown in Fig.~\ref{fig:gates}(a). Naively, when controlled-$U$ is non-Clifford, its constant-depth implementation requires two rounds of feedforward -- one for \textsc{distribute} and one for \textsc{collapse}. However, this can always be reduced to a single round by exploiting the freedom in how the Pauli-$X$ defects in \textsc{distribute} are corrected. Rather than sweeping symmetrically from the middle to each boundary as in Fig.~\ref{fig:overview}(a), one can choose the GHZ qubit used as a control as the ``reference'', i.e., the qubit that requires no correction. This ensures that the feedforward layer commutes with controlled-$U$ and can thus be deferred until \textsc{collapse}.

For an experimental demonstration of an error-detected long-range \textsc{cnot} -- a special case of this primitive -- see Sec.~\ref{ssec:lrcx}.

\subsubsection{Fan-out}
As a second example, we consider the fan-out gate, one of the prototypical examples of a multi-qubit operation implementable in constant depth with dynamic circuits~\cite{Buhrman_StatePreparation_2024, Baumer_EfficientLongRange_2024}. As noted in the introduction, fan-out underlies the constant-depth realization of the $N$-qubit Toffoli gate and the quantum Fourier transform~\cite{Hoyer_QuantumFanout_2005}, and features prominently in a range of recently explored applications of dynamic circuits~\cite{Lund_ConstantDepthQuantum_2026, Yeo_ReducingCircuit_2025, Cao_MeasurementDrivenQuantum_2026}. Its practical importance thus makes its error-detected variant a compelling target\footnote{We note that an error-detected fan-out was previously proposed in Ref.~\cite{Liao_AchievingComputational_2025} using the unitary entangle-disentangle protocol discussed in the introduction. That construction sacrifices constant depth; here we instead pursue an implementation that is both constant-depth and error detected.}.

Our implementation is shown in Fig.~\ref{fig:gates}(b). Because the controlled-$U$ layer is Clifford, the feedforward stage of \textsc{distribute} can be deferred until \textsc{collapse}  such that only one round is needed. Furthermore, both rounds of mid-circuit measurements can be parallelized at the expense of either constant gate overhead or by leveraging increased connectivity. In either case, the resulting circuit inherits the full error-detection capability of \textsc{collapse}$^{(n)}$, flagging both the fusion measurement errors incurred during \textsc{distribute} and the readout errors incurred during \textsc{collapse}$^{(n)}$.

\subsubsection{Multi-qubit Pauli rotation}
Another useful error-detected primitive realizable in constant depth is the class of multi-qubit Pauli rotations,
\begin{equation}
    U = e^{-i\phi/2\bigotimes_{j=1}^N P_j},
\end{equation}
where $P_j\in\{I,X,Y,Z\}$. Our implementation, shown in Fig.~\ref{fig:gates}(c), generalizes the protocol of Ref.~\cite{Yang_HarnessingPower_2024}, here augmented with error detection. Because the controlled-$U$ layer is Clifford, the feedforward stage of \textsc{distribute} can be deferred until \textsc{collapse}$^{(n)}$; similar to fan-out, the mid-circuit measurements themselves can also be deferred at the cost of either a constant gate overhead or increased connectivity. The rotation angle is injected via application of $R_x(\phi)$ to the collapsed ancilla, which is subsequently measured to determine the Pauli corrections on the data qubits. Consequently, when this final rotation is non-Clifford, two rounds of mid-circuit measurement and feedforward are needed to implement a generic multi-qubit Pauli rotation (one preceding the non-Clifford $R_x(\phi)$, and another preceding the first non-Clifford gates that follow the multi-qubit Pauli rotation).

\subsubsection{Hadamard test}
Our final example of a multi-qubit operation is the Hadamard test, an essential primitive across quantum algorithms. Given an $N$-qubit unitary $U$, the Hadamard test estimates the real (or imaginary) part of $\langle \psi|U|\psi \rangle$ by applying $U$ controlled on an ancilla prepared in $\ket{+}$ and measuring in the $X$ (or $Y$) basis. It underlies applications ranging from phase estimation~\cite{Kitaev_QuantumMeasurements_1995} and entanglement spectroscopy~\cite{Johri_EntanglementSpectroscopy_2017} to the measurement of dynamical correlation and response functions~\cite{Somma_SimulatingPhysical_2002, Roggero_DynamicLinear_2019}. Recently, the multivariate  \textsc{SWAP} test -- a special case of the Hadamard test -- was shown to have a constant-depth dynamic-circuit implementation~\cite{Quek_MultivariateTrace_2024}. Our protocol can be viewed as a generalization to arbitrary controlled-$U$ additionally endowed with built-in error detection, thereby opening the technique to a wide range of applications while simultaneously providing a path to ``upgrade'' existing protocols.

The primary cost to implementing a Hadamard test is the depth required to realize the controlled-$U$ operation, which is generally global. For instance, the Hadamard test of a $k$-layer, Trotterized time-evolution operation $U = e^{-iHt}$ -- a common target in phase estimation and other time-series algorithms~\cite{Somma_QuantumEigenvalue_2019, Parrish_QuantumFilter_2019, Shen_EstimatingEigenenergies_2025} -- requires depth $O(Nk)$ in the worst case. This cost has motivated a variety of algorithmic and statistical techniques that reduce or eliminate the controlled operation, ranging from the use of GHZ-like reference states to schemes that resolve only energy gaps rather than the full spectrum~\cite{Patel_QuantumPhaselift_2026}. Each involves its own tradeoffs; our aim here is instead to reduce the cost of the controlled operation directly.

\begin{figure*}
    \centering
    \includegraphics[width=1\linewidth]{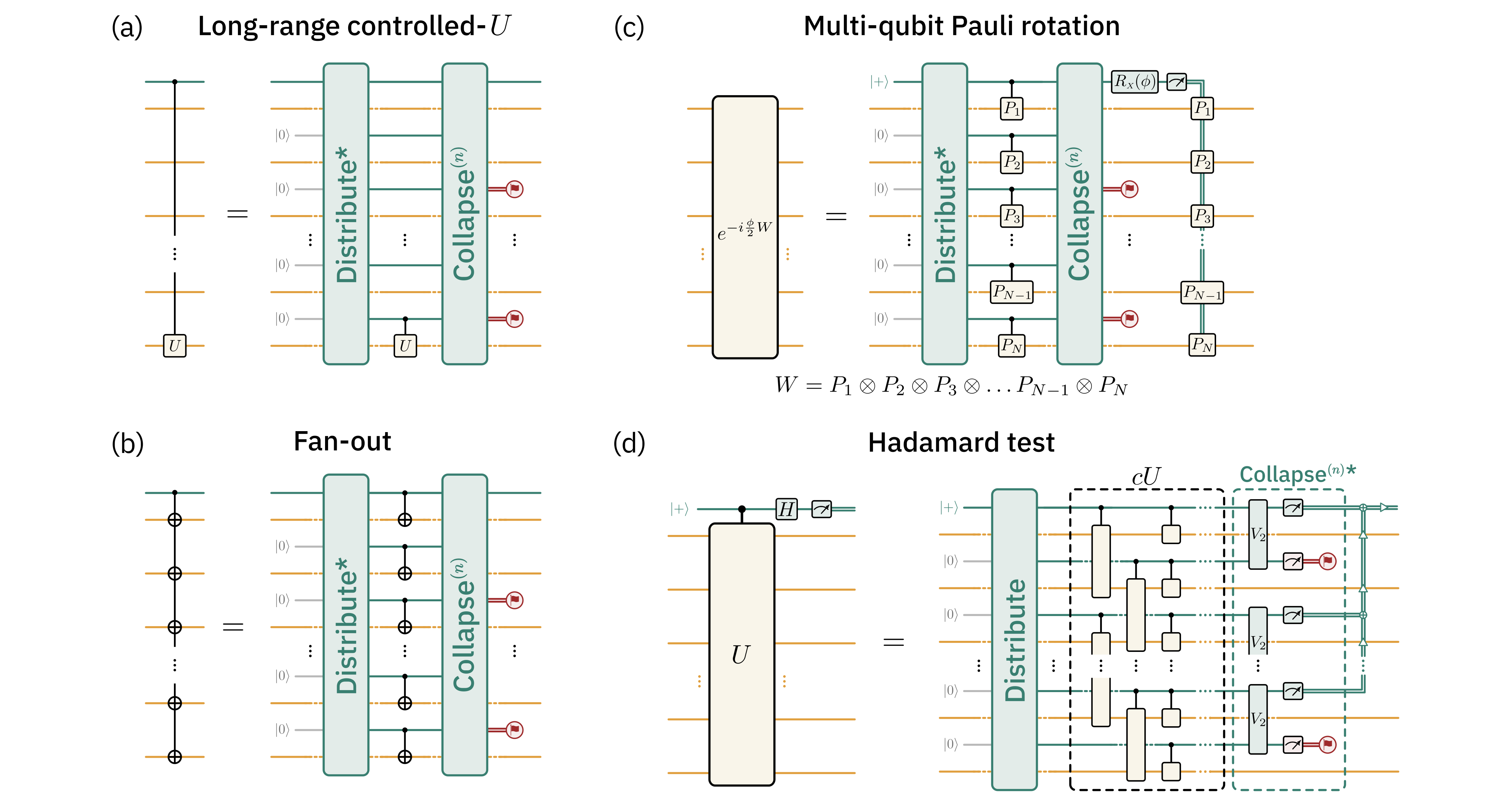}
    \caption{Examples of constant-depth, long-range and multi-qubit primitives augmented with error-detection using the \textsc{distribute}-\textsc{collapse} framework. 
     \textsc{distribute}$^*$ denotes a variant of \textsc{distribute} that defers feedforward corrections to the \textsc{collapse} stage of the circuit, while  \textsc{collapse}$^{(n)*}$ indicates the feedforward corrections can be carried out in post-processing. (a) A long-range two-qubit controlled-unitary. (b) An error-detected, constant-depth fan-out gate. (c)  An error-detected, constant-depth multi-qubit Pauli rotation. (d) An error-detected, depth $O(k)$ implementation of the Hadamard test for an arbitrary unitary $U$ of depth $k$, requiring a single round of mid-circuit measurements and feedforward. For specificity, we show the Hadamard test for estimating  $\mathrm{Re}\,\langle U \rangle$; $\mathrm{Im}\,\langle U \rangle$ can be estimated in a similar fashion by swapping the initial ancilla state $|+\rangle$ with $|-i\rangle$.}
    \label{fig:gates}
\end{figure*}

Our implementation is shown in Fig.~\ref{fig:gates}(d). For the example of a Hadamard test for a time evolution operator with $k$ Trotter steps, distribution of the control enables an $O(k)$ depth realization. Moreover, because the final $X$ (or $Y$) basis measurement of the Hadamard test is Clifford, the feedforward corrections of \textsc{collapse}$^{(n)}$ need not be applied prior; instead, this final measurement can be carried out in parallel with the $X$-basis readout of \textsc{collapse}$^{(n)}$ and the associated feedforward $Z$ correction applied in post-processing. Equivalently, at the operator level, one can view this as extracting the desired phase by directly measuring the joint observables $X_1 X_2 X_3\ldots X_N$ (real part) or $Y_1 X_2 X_3\ldots X_N$ (imaginary part) across the full distributed $N$-qubit GHZ state. In either view, the entire protocol requires only the single round of mid-circuit measurement and feedforward in \textsc{distribute} for general controlled-$U$; if it is Clifford, the feedforward can be eliminated entirely, with the corrections applied in post-processing.

Crucially, error detection is built in at no additional feedforward cost: the same measurements that recover the phase encoded in the distributed control simultaneously flag fusion, terminal-measurement, and ancilla bit-flip errors throughout the Hadamard test. Altogether, the net result is a constant-depth, error-detected Hadamard test that we regard as one of the central results of this work -- a concrete and broadly applicable use case for dynamic circuits that, in contrast to the largely bespoke state preparation and gate synthesis protocols explored in the literature to date, delivers a genuine depth advantage across a wide range of quantum algorithms, all while incorporating a low-overhead error detection strategy.

\subsection{State preparation}
\label{ssec:states}
Constant-depth preparation of states with long-range correlations is one of the most promising applications of dynamic circuits~\cite{Tantivasadakarn_LongRangeEntanglement_2024, Smith_ConstantDepthPreparation_2024, Buhrman_StatePreparation_2024, Piroli_ApproximatingManyBody_2024}. In this Section, we provide several examples of dynamic-circuit state preparation protocols that can be expressed in our \textsc{distribute}--\textsc{collapse} framework and thus admit a low-overhead error detection strategy. As in the prior section, these examples are not meant to be exhaustive, but rather serve as illustrative examples where our error detection scheme can potentially boost the robustness of existing protocols for preparing broad classes of states.

\subsubsection{GHZ state}
\label{sssec:ghz}

The GHZ state provides the simplest example of a state preparation protocol that admits integration of our error detection scheme. The steps are as follows: (i) use \textsc{distribute$^*$} on register $A$; (ii) grow the GHZ state onto register $B$ using a single layer of \textsc{cnot} gates, with registers $A$ and $B$ as the control and target, respectively; (iii) implement \textsc{collapse}$^{(n)}$ on register $A$, leaving a GHZ state on register $B$\footnote{Identically, this procedure can be understood as carrying out an error-detected fan-out gate on control ($A$) and data ($B$) registers initialized to $|+\rangle$ and $|000\ldots 0\rangle$, respectively}. This routine effectively consumes one GHZ state to prepare a second that, depending on the details of the hardware noise, can be of improved overall fidelity. However, because our error detection scheme only detects bit-flip errors, it will generally improve GHZ populations at the expense of degraded coherence relative to the initial GHZ state. 

A second complementary strategy is to lift \textsc{distribute} to higher dimensions where redundancy can be built into the fusion measurements themselves. As this approach falls outside of the main focus of this work, we provide further details in App.~\ref{app:GHZ}.

\subsubsection{W states, Dicke states, and towers of excited states}
\label{sssec:w_and_dicke}

\begin{figure}
    \centering
    \includegraphics[width=1\linewidth]{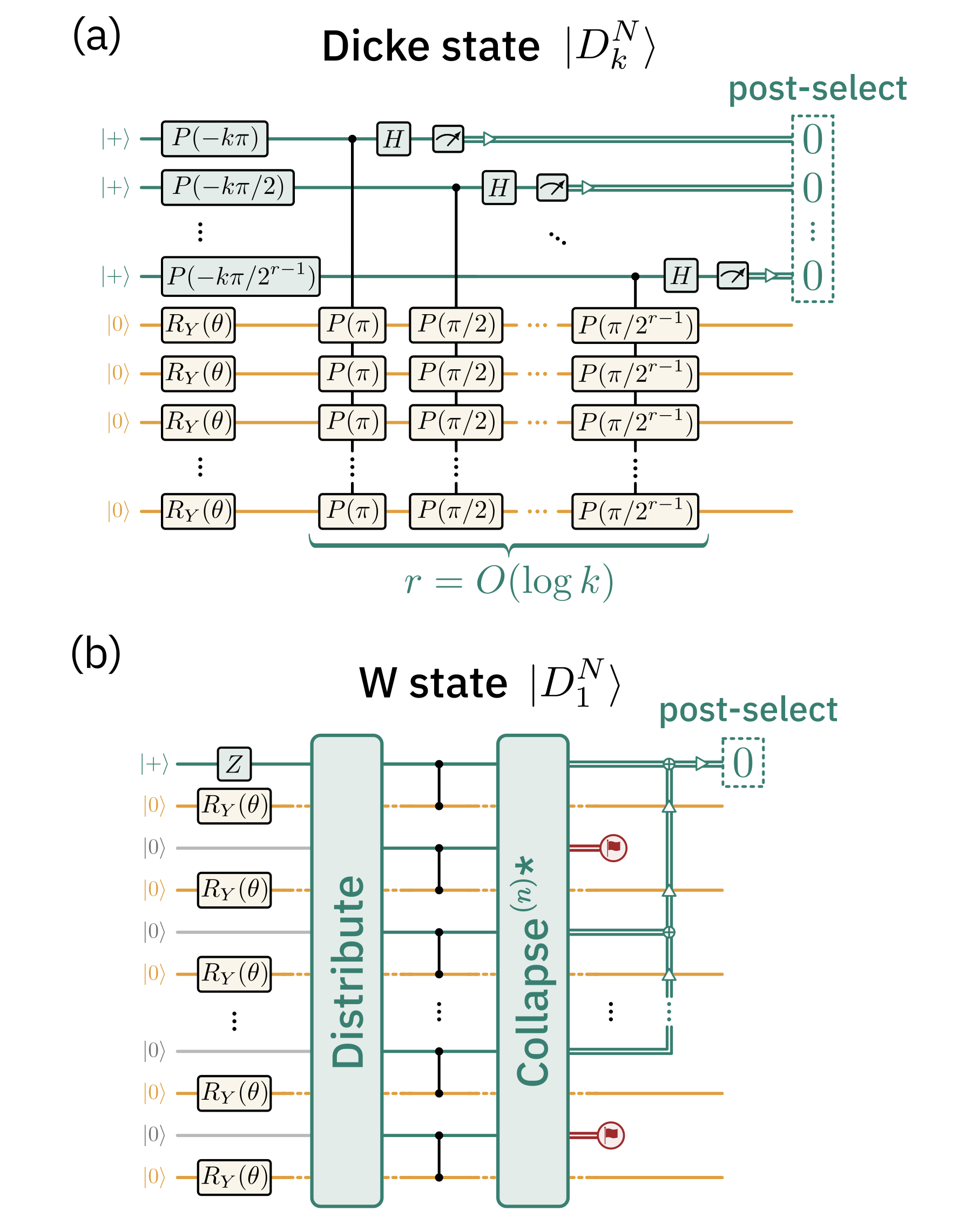}
    \caption{Error-detected preparation of Dicke states. (a) A simplified variant of the protocol of Ref.~\cite{Piroli_ApproximatingManyBody_2024} for approximately preparing the $N$-qubit Dicke state of weight $k$, $|D_k^N\rangle$. Single-qubit $R_Y(\theta)$ rotations first prepare a binomially-weighted product state, which is then projected onto the target sector $k$ via $r=O(\log k)$ rounds of bit extraction, each a layer of controlled-phase gates followed by an ancilla measurement postselected on 0. Each round can be realized by a single round of \textsc{distribute}--\textsc{collapse} and augmented with error detection, in analogy to the Hadamard test of Fig.~\ref{fig:gates}(d). (b) Constant-depth, error-detected preparation of the W state $|D^N_1\rangle$, requiring one round of mid-circuit measurement and feedforward.}
    \label{fig:stateprep}
\end{figure}

Dicke states $\{\ket{D^N_k}\}$, the symmetric superpositions of all weight-$k$ bitstrings on $N$ qubits, constitute an essential class of long-range entangled states for which efficient dynamic-circuit preparation protocols have been developed~\cite{Buhrman_StatePreparation_2024, Piroli_ApproximatingManyBody_2024,  Yu_EfficientPreparation_2026}. Beyond their fundamental interest, Dicke states are a valuable resource for quantum metrology~\cite{Sorensen_EntanglementExtreme_2001, Toth_MultipartiteEntanglement_2012, Lucke_TwinMatter_2011} and constrained quantum optimization~\cite{Hadfield_QuantumApproximate_2019, Bartschi_GroverMixers_2020}, and serve as useful initial states for preparing nontrivial many-body entanglement~\cite{VanDyke_PreparingBethe_2021, Guo_MeasurementBasedPreparation_2026a}. Moreover, efficient protocols to prepare their simplest nontrivial instance $\ket{D^N_1}$, the W state, have enabled recent studies of particle scattering and energy-resolved transport on quantum processors~\cite{Farrell_DigitalQuantum_2025, Lee_StudyingEnergyresolved_2026}. We now describe how the preparation of both W states and more general Dicke states can be augmented with our error detection scheme with no additional ancillas and only a constant-depth gate overhead.

The global symmetry of Dicke states can be leveraged to realize the remarkably simple preparation strategy shown in Fig.~\ref{fig:stateprep}(a): using only single-qubit rotations, first prepare a product state whose expansion in the Dicke basis is binomially distributed across the different weight sectors; following this, project onto the target sector of definite weight $k$. This latter step amounts to the measurement (and postselection) of the global observable $\hat{n} = \sum_j{(1-Z_j)/2}$ which, as shown in Ref.~\cite{Piroli_ApproximatingManyBody_2024}, can be performed in constant depth using dynamic circuits. While this protocol is probabilistic, it succeeds with probability $O(1/\sqrt{k})$ -- independent of system size for $k \ll N$, and degrading only to $O(1/\sqrt{N})$ in the worst case of half-filling $k \approx N/2$. This success probability can either be accepted directly or boosted via amplitude amplification. Moreover, while $O(\log N)$ bits of precision are required for exact preparation, only $O(\log k)$ are needed for  faithful approximate preparation due to the fast-decaying tails of the initial binomial distribution. See Appendix~\ref{app:dicke} for details.

The global measurement underlying this protocol, shown for general $k$ in Fig.~\ref{fig:stateprep}(a), is naturally expressed within the \textsc{distribute}--\textsc{collapse} framework and can therefore be augmented with error detection. Fig.~\ref{fig:stateprep}(b) shows an explicit example for the W state ($k=1$), which requires just one round of mid-circuit measurement and feedforward. In the general case, each round of bit extraction (here reduced to a Hadamard test with a layer of controlled phase gates) is replaced by a round of \textsc{distribute}--\textsc{collapse}. As shown in Ref.~\cite{Piroli_ApproximatingManyBody_2024} and discussed further in App.~\ref{app:dicke}, the full dynamic circuit can be implemented in various ways with distinct spacetime tradeoffs. In particular, the $O(\log k)$ bits of $\hat{n}$ can be extracted either (i) in parallel, via $O(\log k)$ simultaneous instances of \textsc{distribute}--\textsc{collapse} acting on $O(N\log k)$ total ancillas in $O(1)$ depth; or (ii) sequentially, using $O(\log k)$ rounds of \textsc{distribute}--\textsc{collapse} on the same $O(N)$ ancillas, similar to iterative phase estimation. Because the target sector is fixed, this sequential readout requires no adaptivity between rounds.

Finally, we note that the above strategy for preparing Dicke states has been extended~\cite{Guo_TowerStructured_2026} to towers of many-body scar states of the form
\begin{equation}
|E_k\rangle = \frac{(\mathcal{J}^{\dagger})^k}{\mathcal{N}_k}|\Psi_0\rangle,
\end{equation}
where $|\Psi_0\rangle$ is an efficiently preparable initial state, $\mathcal{N}_k$ a normalization factor, and $\mathcal{J}^{\dagger}$ a creation operator for an excitation of momentum $q$,
\begin{equation}
\mathcal{J}^{\dagger} = \sum_{j= 1}^N e^{iqj}O_j^{\dagger},
\end{equation}
with $O^{\dagger}_j$ a local excitation operator with finite support. Similar to above, the core idea is to use quantum phase estimation to project an easily prepared state onto a long-range entangled state of definite excitation number. Our error-detection scheme therefore incorporates naturally 
into this broader family of protocols.

For an experimental demonstration of the error-detected preparation of the W state, see Sec.~\ref{ssec:wstate}.

\subsubsection{Non-normal matrix product states}

As a final example, we turn to the protocol introduced in Ref.~\cite{Smith_ConstantDepthPreparation_2024} for preparing non-normal matrix product states -- a class of entangled many-body states that characterize the degenerate ground states of gapped local Hamiltonians in 1D and exhibit long-range correlations. Physically, they are closely related to symmetry-breaking phases of matter~\cite{Schuch_ClassifyingQuantum_2011}. For the purpose of the present discussion, we put aside these technical considerations and instead view them simply as GHZ-like superpositions of the form
\begin{equation}
|\Psi\rangle = \frac{1}{\mathcal{N}}\sum_{k = 0}^{K-1} |\psi_{\textrm{MPS}, k}\rangle,
\end{equation}
where each constituent $|\psi_{\textrm{MPS}, k}\rangle$ is a \emph{normal} matrix product state -- the class characterizing the unique (non-degenerate) ground states of gapped local Hamiltonians in 1D, exhibiting only short-range correlations.

We focus on the case $K=2$ as it is most naturally expressed in terms of the framework introduced here; we expect that $K>2$ will require a generalization of our error detection strategy, and we leave it as an avenue for future research. The core preparation strategy, here cast in terms of \textsc{distribute} and \textsc{collapse}, is simple: 

\begin{enumerate}
    \item First, prepare an ancillary distributed GHZ state of the form $|\Psi_0\rangle =(1/\mathcal{N})\sum_{k=0}^1|k\rangle^{\otimes N}\otimes|0\rangle^{\otimes N}$ using \textsc{distribute}.
    
    \item Next, within each sector $k$, conditionally prepare the normal matrix product state $|\psi_{\textrm{MPS}, k}\rangle$, yielding the state $|\Psi_0'\rangle = (1/\mathcal{N})\sum_{k=0}^1|k\rangle^{\otimes N}\otimes|\psi_{\textrm{MPS}, k}\rangle$. If $|\psi_{\textrm{MPS}, k}\rangle$ obeys certain conditions, this step is implementable in constant depth with dynamic circuits (see Ref.~\cite{Smith_ConstantDepthPreparation_2024} for details); otherwise, log-depth methods, applicable to arbitrary normal matrix product states, can be substituted~\cite{Malz_PreparationMatrix_2024, Murota_ExactLogdepth_2026}.
    
    \item Finally, fully measure out all GHZ qubits via \textsc{collapse}$^{(n)*}$; an even parity outcome, which has probability 1/2, heralds successful preparation of the target state $|\Psi\rangle$. 
    
    \end{enumerate}
Cast into this form, incorporation of error detection is immediate: one simply replaces \textsc{collapse}$^*$ with its error-detected counterpart \textsc{collapse}$^{(n)*}$ for $n\geq 2$, thus augmenting preparation protocols for an essential class of states with low-overhead error detection. 

As a final note, we emphasize that the above protocol, while here framed in the context of non-normal matrix product states, applies to the preparation of superpositions of any efficiently preparable states heralded at constant probability. In this sense it is less a single example than a general recipe: any protocol that prepares a state by collapsing a GHZ-like superposition inherits our error detection essentially for free, with no additional ancillas and only a constant-depth gate overhead.

\section{Experimental results}
\label{sec:experiment}

We demonstrate the immediate utility of our error-detected framework through a pair of experiments carried out on a Heron r3 quantum processor, \texttt{ibm\_boston}. First, we demonstrate an error-detected long-range \textsc{cnot} gate and use it to prepare Bell pairs separated by up to $\ell = 100$ qubits, certifying entanglement via fidelity to the ideal Bell state. At this largest separation, we report a post-selected fidelity of $F=0.59\pm 0.02$ with error-detection, comfortably above the $F > 0.5$ entanglement threshold and a significant improvement over the non-error-detected baseline, $F=0.39 \pm 0.01$ (Sec.~\ref{ssec:lrcx}). Following this, we prepare up to $N=20$ qubit W states using the constant-depth protocol of Ref.~\cite{Piroli_ApproximatingManyBody_2024}, here upgraded with error detection, leading to absolute fidelity improvements of $\Delta F \approx 0.2$ over the baseline implementation (Sec.~\ref{ssec:wstate}). The sections below elaborate on our experimental protocols and findings.

\begin{figure*}
    \centering
    \includegraphics[width=1\linewidth]{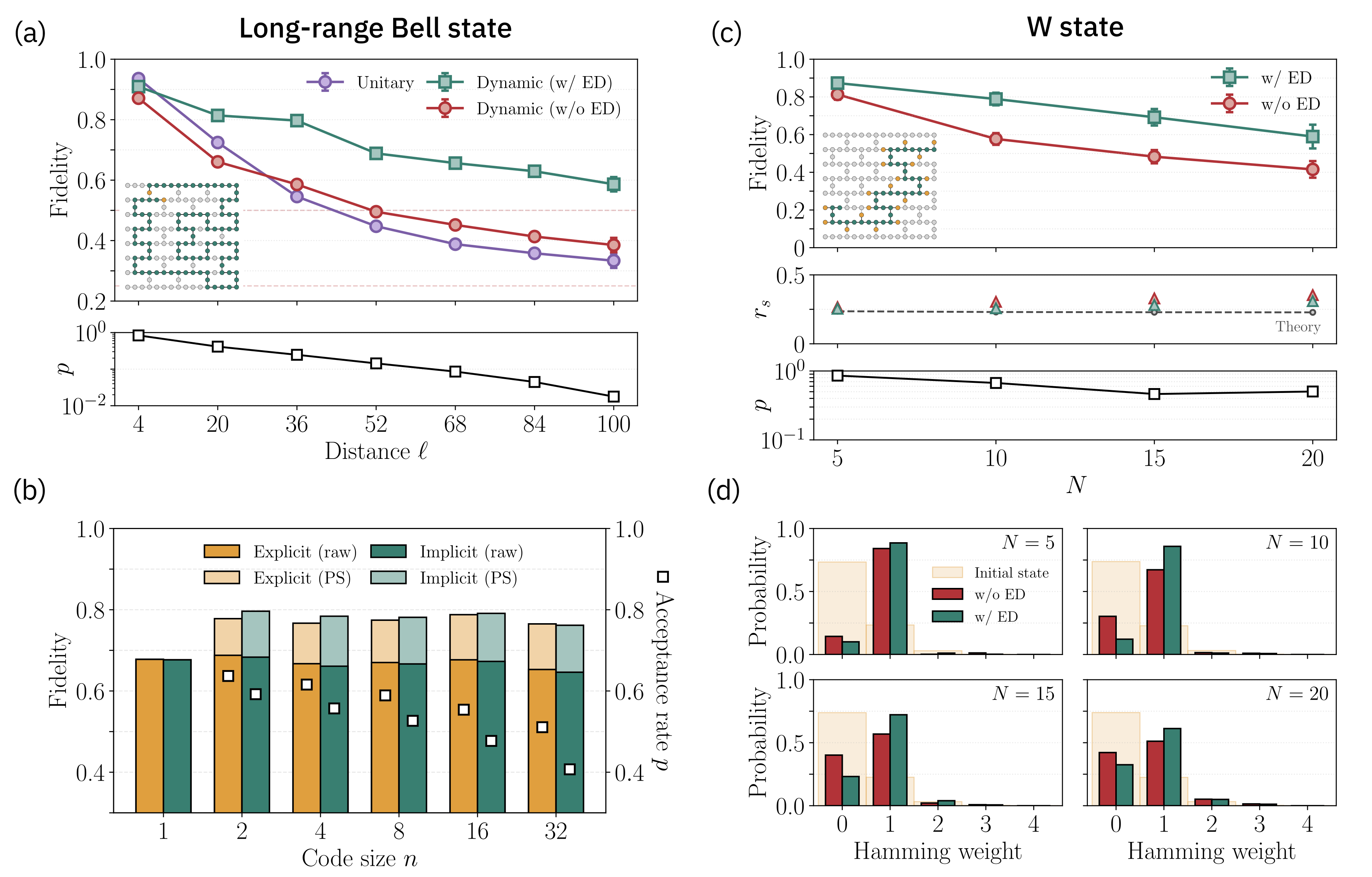}
    \caption{Experimental demonstration of our error detection framework on \texttt{ibm\_boston}. Left: error-detected preparation of a long-range entangled Bell state $|\Phi^+\rangle = (|00\rangle + |11\rangle)/\sqrt{2}$ across $\ell$ qubits. (a) Fidelity for three protocols: the $O(1)$-depth dynamic-circuit protocol with (green) and without (red) error detection, and the $O(\ell)$-depth unitary circuit (purple). Error detection uses implicit checks with $n=2$; the corresponding acceptance rate $p$ is shown in the lower panel. The upper and lower dotted red lines indicate the certified-entanglement threshold ($F>0.5$) and the fidelity of a maximally mixed state ($F=0.25$), respectively.   (b) For fixed $\ell = 32$, the fidelity (bars) and acceptance rate (square markers) for explicit (yellow) and implicit (green) check types at distinct code sizes $n$. Fidelities are reported before (raw) and after (PS) postselection. Right: error-detected preparation of a W state. (c) Fidelity to the ideal W state with (green) and without (red) error detection. The error-detected case uses $n=2$ implicit checks, with the corresponding acceptance rate $p$ shown in the lower panel. The middle panel shows the projection success rate $r_s$ for each case, compared with the theoretical expectation (dashed gray line). (d) Hamming weight distribution of the prepared state for each size $N$; for reference, the distribution of the initial product state before projection is shown in pale yellow.}
    \label{fig:experiment}
\end{figure*}

\subsection{Long-range CNOT}
\label{ssec:lrcx}

We first turn to the implementation of a long-range \textsc{cnot}, a special case of the long-range controlled-$U$ in Fig.~\ref{fig:gates}(a) and a canonical benchmark for dynamic-circuit protocols~\cite{Baumer_EfficientLongRange_2024, Niu_ACDC_2024, Hashim_EfficientGeneration_2025, Liao_AchievingComputational_2025}. In particular, we use it to prepare a long-range entangled Bell pair separated by $\ell$ qubits, using fidelity relative to the ideal Bell state $|\Phi^+\rangle = (|00\rangle + |11\rangle)/\sqrt{2}$ to assess performance and certify entanglement. As discussed in Sec.~\ref{ssec:gates}, the circuit structure for a long-range controlled-$U$ gate allows \textsc{distribute} to be replaced with its feedforward-deferred counterpart such that only one round of mid-circuit measurement and feedforward is needed. For explicit circuit diagrams, see App.~\ref{app:lrcx}.

We report our findings in Fig.~\ref{fig:experiment}. Panel (a) shows the fidelity $F = \left(1 + \langle XX\rangle - \langle YY \rangle + \langle ZZ\rangle \right)/4$
as a function of separation $\ell$ (with $\ell + 2$ qubits in total) for three implementations: the constant-depth dynamic circuit with error detection ($n=2$, implicit checks), the same circuit without error detection ($n=1$), and the static unitary approach that requires a circuit of depth $O(\ell)$ (see App.~\ref{app:lrcx}). We observe that, without error detection, the dynamic circuit confers no advantage over its unitary counterpart until a ``crossover'' between $\ell = 20$ and $\ell = 36$, above which it yields only limited fidelity gains of $\sim 0.05$. This modest improvement reflects the cost of the $\ell/2$ mid-circuit measurements and associated feedforward latency, which together limit the realized benefit of the reduced depth. 

In contrast, the error-detected dynamic circuit dramatically improves upon the unitary protocol: the crossover occurs at a smaller system size, and an absolute fidelity gain of $\sim 0.25$ over the unitary protocol is observed for all $\ell \geq 36$. At the longest distance studied, $\ell = 100$, our error-detected protocol achieves $F = 0.59 \pm 0.02$, comfortably surpassing the threshold for certified entanglement ($F > 0.5$, top dotted red line). Neither alternative clears this threshold, as the non-error-detected dynamic and unitary circuits reach only $F = 0.39 \pm 0.01$ and $F = 0.33 \pm 0.01$, respectively, experimentally validating the promise of our protocol. While this improvement comes at the cost of a postselection overhead, it remains modest even at $\ell = 100$ with an acceptance rate $p\gtrsim 10^{-2}$.

Separately, we investigate the dependence of the fidelity on the code size $n$ and check type for fixed $\ell = 32$. As shown in Fig.~\ref{fig:experiment}(b), increasing the code size beyond $n=2$ does not substantively improve the fidelity, and comes at the cost of a monotonically decreasing acceptance rate. For small $n$, implicit checks generally outperform explicit checks, while at large $n$, the additional gate and depth overhead of implicit checks outweighs their improved detection capability. Altogether, we find that $n=2$ with implicit checks offers a good middle ground, capturing most of the benefit of error detection while minimizing the postselection overhead, hence our use of this combination for both the distance-sweep experiment of panel (a) and the W state experiments which we now discuss.

\subsection{W state preparation}
\label{ssec:wstate}

In the above example, the long-range \textsc{cnot} manifests as a simplified instance of our general framework, involving a controlled-$U$ operation targeting just a single qubit. For our second demonstration, we extend to the genuine multi-qubit setting; in particular we carry out the error-detected, constant-depth preparation of the W state: the symmetric superposition of all Hamming weight one bitstrings over $N$ qubits, 
\begin{equation}
    |W\rangle \equiv |D_1^N\rangle = \frac{1}{\sqrt{N}}\sum_{j=1}^{N}|w_j\rangle,
\end{equation}
where $|w_j\rangle =|0\rangle^{\otimes j-1}\otimes |1\rangle\otimes |0\rangle^{\otimes N-j}$ denotes a state with a single excitation at site $j$ and zeros elsewhere. 

We carry out the protocol summarized in Sec.~\ref{sssec:w_and_dicke} and detailed in App.~\ref{app:dicke}. The circuit diagram is illustrated in Fig.~\ref{fig:stateprep}(b). For the initial product state distribution, we use the single-qubit rotation angle $\theta = -2\sin^{-1}(\sqrt{0.3/N})$, chosen to balance the theoretically expected projection success rate $r_s$ with fidelity; in the limit $N\to\infty$, this yields $r_s \approx 0.226$ and $F_{\textrm{theory}} \approx 0.986$. We prepare W states of size $N = 5,\,10,\,15$ and $20$ on \texttt{ibm\_boston} using the layout shown in the inset of Fig.~\ref{fig:experiment}(c), consisting of a 1D line of ancillas with ``dangling'' data qubits, requiring $3N$ total qubits. 

Our findings are shown on the right side of Fig.~\ref{fig:experiment}. In panel (c), we report the fidelity with ($n=2$, implicit checks) and without ($n=1$) error detection, obtained via direct fidelity estimation~\cite{Flammia_DirectFidelity_2011} using 50{,}000 Pauli samples; at the largest size $N=20$, this requires 381 unique measurement bases, each allocated a number of shots determined from observed postselection rates. Error bars are estimated via bootstrapping. See App.~\ref{app:wstate} for further details. 

We find that error detection substantially improves the fidelity at all sizes, reaching $F = 0.87\pm0.03$, $0.79\pm0.03$, $0.69\pm0.04$, and $0.59\pm0.06$ for $N = 5$, $10$, $15$, and $20$, respectively, corresponding to absolute gains of $\Delta F = 0.06\pm 0.04$, $0.21 \pm 0.04$, $0.21 \pm 0.05$, and $0.17 \pm 0.08$ over the baseline without error detection. We further observe closer agreement between the measured and theoretically predicted projection success rate, indicating that error detection improves the quality of the projection. This improvement comes at the cost of a modest postselection overhead, with the circuit-averaged acceptance rate $p$ ranging from $0.86$ to $0.46$.

To further characterize the quality of the prepared W states, we examine the Hamming weight distribution of the prepared state in the computational basis -- see Fig.~\ref{fig:experiment}(d). For an exact preparation we expect weight one with unit probability; because the executed protocol is approximate, however, small contributions from higher odd Hamming weights are expected. Without error detection, the observed distribution shows significant leakage to weight zero. This reflects an imperfect global odd-parity projection: a phase error in the ancillary GHZ state anticommutes with the all-$X$ observable measured during \textsc{collapse}, flipping the inferred parity and causing the protocol to accept a state projected on the even-parity sector, of which the all-zeros state is the dominant contribution. With error detection, the projection quality improves, as evidenced by the enhanced weight-one peak and the suppression of the all-zeros component across all sizes. A non-negligible all-zeros component nonetheless persists, arising from coherence loss of the ancillary GHZ state due to  undetectable errors, such as dephasing and crosstalk. Improving the performance of this and similar dynamic-circuit protocols will therefore require dedicated strategies for suppressing such errors.

\section{Conclusion}
\label{sec:conclusion}

By combining unitary gates with mid-circuit measurements and classical feedforward, dynamic circuits enable the constant-depth implementation of a range of many-qubit unitaries and state preparations that, with static unitary circuits alone, provably require a depth scaling with system size. This depth reduction, however, is not free: it typically requires a number of mid-circuit measurements that grows with system size and which, on present-day devices, introduce errors that degrade the long-range entanglement these protocols rely on. The result is a tension in which, depending on the specifics of the hardware, the near-term promise of dynamic circuits can be undercut by the very measurements that enable it. 

In this work, we have introduced a low-overhead error-detection scheme to resolve this tension, organized around the concept of distributed control and realized through a pair of primitives, \textsc{distribute} and \textsc{collapse}$^{(n)}$. Rather than treating readout errors as an unavoidable cost of dynamic-circuit protocols, we show that the structure of \textsc{collapse}$^{(n)}$ admits lightweight checks that can flag these and other errors.

The value of our scheme is twofold. First, it unifies a broad collection of dynamic-circuit protocols under a single operational framework, illustrated here through a number of explicit examples. These range from the implementation of low-level primitives, such as long-range and many-qubit gates, to the preparation of long-range entangled states, including the W state, higher-weight Dicke states, and non-normal matrix product states. We also extend our framework to higher-level algorithmic primitives, providing a low-depth implementation of the Hadamard test as a flagship example.  Second, our framework offers a simple, low-overhead means to ``upgrade'' each of these protocols with a common error-detection scheme. This upgrade requires no additional ancillas and incurs only a constant-depth gate cost, ultimately trading classical postselection overhead for improved fidelity.

To illustrate the practical relevance of our scheme for near-term applications, we presented a pair of experimental demonstrations carried out on IBM Quantum hardware. In the first, we implemented a long-range \textsc{cnot} gate to prepare an entangled Bell pair separated by as many as 100 intermediate qubits, finding that our error detection scheme enables the (postselected) preparation of certified entanglement across this separation with fidelity $F = 0.59\pm0.02$ and acceptance rate $p > 10^{-2}$; in contrast, neither the unitary counterpart nor the dynamic circuit without error detection meet this threshold, reaching fidelities of $F = 0.39\pm 0.01$ and $F = 0.33\pm 0.01$, respectively. 

In the second demonstration, we implemented the W state preparation protocol of Ref.~\cite{Piroli_ApproximatingManyBody_2024}, here cast into our unified framework. With error detection incorporated, we observed an absolute improvement in fidelity of $\Delta F\approx 0.2$ for W states of up to 20 qubits (requiring up to 60 qubits total counting ancillas), with acceptance rates sustaining $p>0.46$ across all system sizes. Altogether, these findings demonstrate that our error-detection framework can bring the theoretical promise of dynamic circuits closer to present experimental reality.

Looking ahead, while the present work has focused on error detection as it presents an attractive target for near-term experiments, a natural next step is to extend the framework toward active error \emph{correction}. Rather than discarding the flagged shots, one could leverage the check outcomes to correct them. This might range from simply taking a majority vote across the redundant checks within a patch, yielding a more reliable estimate of the required correction, to a more complete decoding strategy that infers and corrects the underlying error; in cases where controlled-$U$ is Clifford, this correction requires no alteration to the circuit structure. We leave a systematic treatment, including extension to non-Clifford controlled-$U$, to future work. More broadly, we anticipate that error detection, and eventually correction, will be an essential ingredient for integration of dynamic circuits into utility-scale applications.

\section*{Acknowledgements}
Contributions by S.M. were primarily completed during an internship at IBM.  K.C.S., B.P. and M.T. are grateful to Ilan Rosen, Alireza Seif, Liran Shirizly, Ioannis Tsioutsios, James Raftery, Jay-U Chung, Kunal Sharma, Will Kirby, Abhinav Kandala, Jake Lishman, Kit Barton, Michael Healy, and Brian Donovan for helpful discussions. 
\appendix

\section{Connectivity considerations: heavy-hex and square lattices}

\begin{figure}[h!]
    \centering
    \includegraphics[width=1\linewidth]{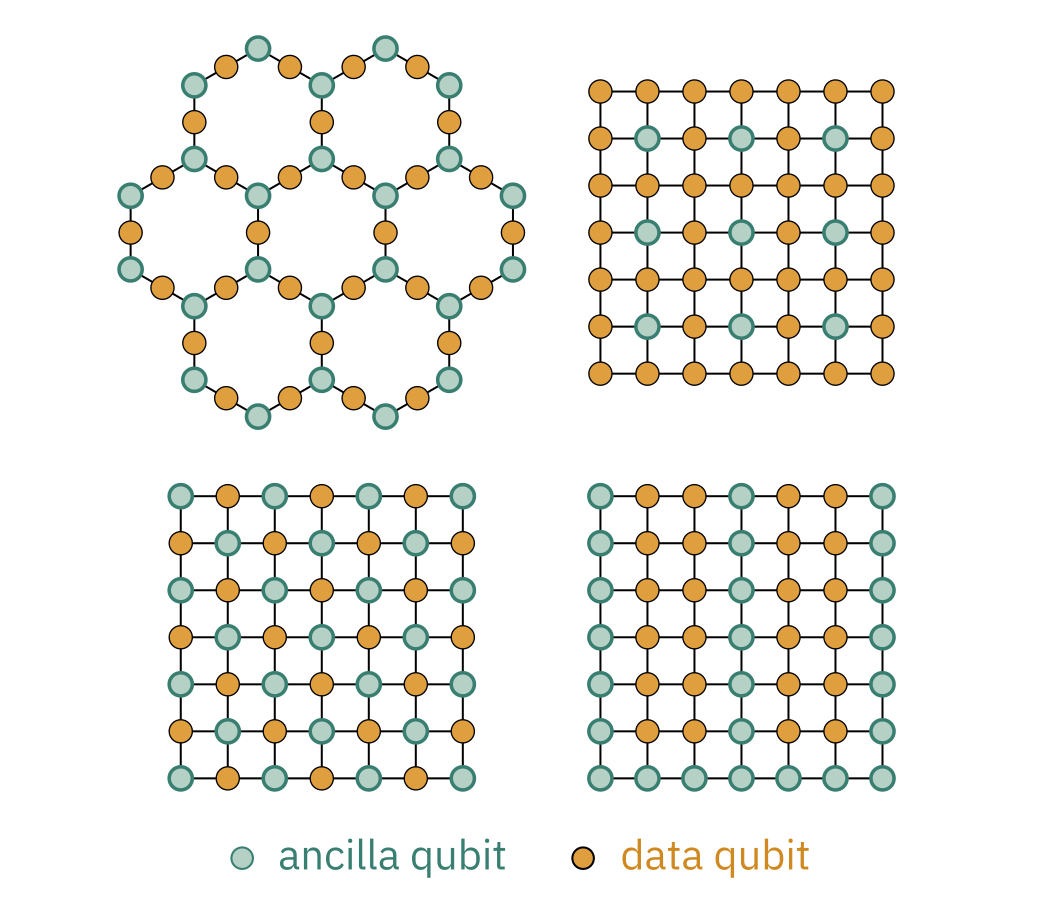}
    \caption{Examples of ancilla-data partitioning on heavy-hex and square lattices. While a 1-to-1 ratio of ancilla and data qubits maximizes parallelization of controlled operations (as in the top- and bottom-left), the ancilla count can be reduced by a constant factor at the expense of a corresponding increase in circuit depth.}
    \label{fig:app_connectivity}
\end{figure}

To enable constant-depth primitives on $N$ data qubits, the  \textsc{distribute}--\textsc{collapse} framework
necessarily requires $O(N)$ ancillas distributed strategically across the system. Although the circuit diagrams throughout this work assume $N$ ancillas, so as to maximally parallelize the controlled gates, we emphasize that this is not strictly necessary: the ancilla count can be reduced by a constant factor at the expense of a corresponding increase in circuit depth by using the same ancilla for multiple controlled operations in sequence. Whether this tradeoff is favorable will depend on architectural details such as connectivity, the availability of fresh qubits, and coherence times, as well as the specific application. For examples of possible data-ancilla partitionings on heavy-hex and square lattices, see Fig.~\ref{fig:app_connectivity}.

\section{Error-detected preparation of GHZ states in higher dimensions}
\label{app:GHZ}

The central idea of this work is to unify a range of common dynamic-circuit primitives under the \textsc{distribute}--\textsc{collapse} framework, and to show that each can be ``upgraded'' simply by replacing \textsc{collapse}$^{(1)}$ with its error-detected variant \textsc{collapse}$^{(n)}$ for $n\geq 2$. As noted in Section~\ref{sssec:ghz}, however, an independent yet complementary error-detection strategy emerges when \textsc{distribute} is lifted to higher dimensions.

To explain, first recall the steps underlying GHZ preparation with dynamic circuits: (i) prepare $k$ independent GHZ ``patches'', each of size $N/k$; (ii) measure $ZZ$ on the $m$ \emph{fusion edges} joining adjacent patches; (iii) reset and re-entangle the measured fusion qubits (those encoding each $ZZ$); (iv) correct the state via feedforward, applying Pauli-$X$ gates conditioned on the parity of an appropriate subset of the mid-circuit measurement outcomes. Intuitively, these corrections ``sweep'' the outcome-dependent domain walls out to the graph boundary where they can be removed. For reference, see Fig.~\ref{fig:overview}(a) which carries out this protocol (taking the control to be $|+\rangle$) on a 1D graph for $k=N/2$. Although $k$ can be chosen freely -- trading circuit depth for mid-circuit measurements and vice versa -- in 1D the total number of fusion edges is fixed at $m = k-1$, so that all $ZZ$ measurement outcomes are statistically independent. As a consequence, there is no redundancy to leverage for error detection.

\begin{figure}
    \centering
    \includegraphics[width=1\linewidth]{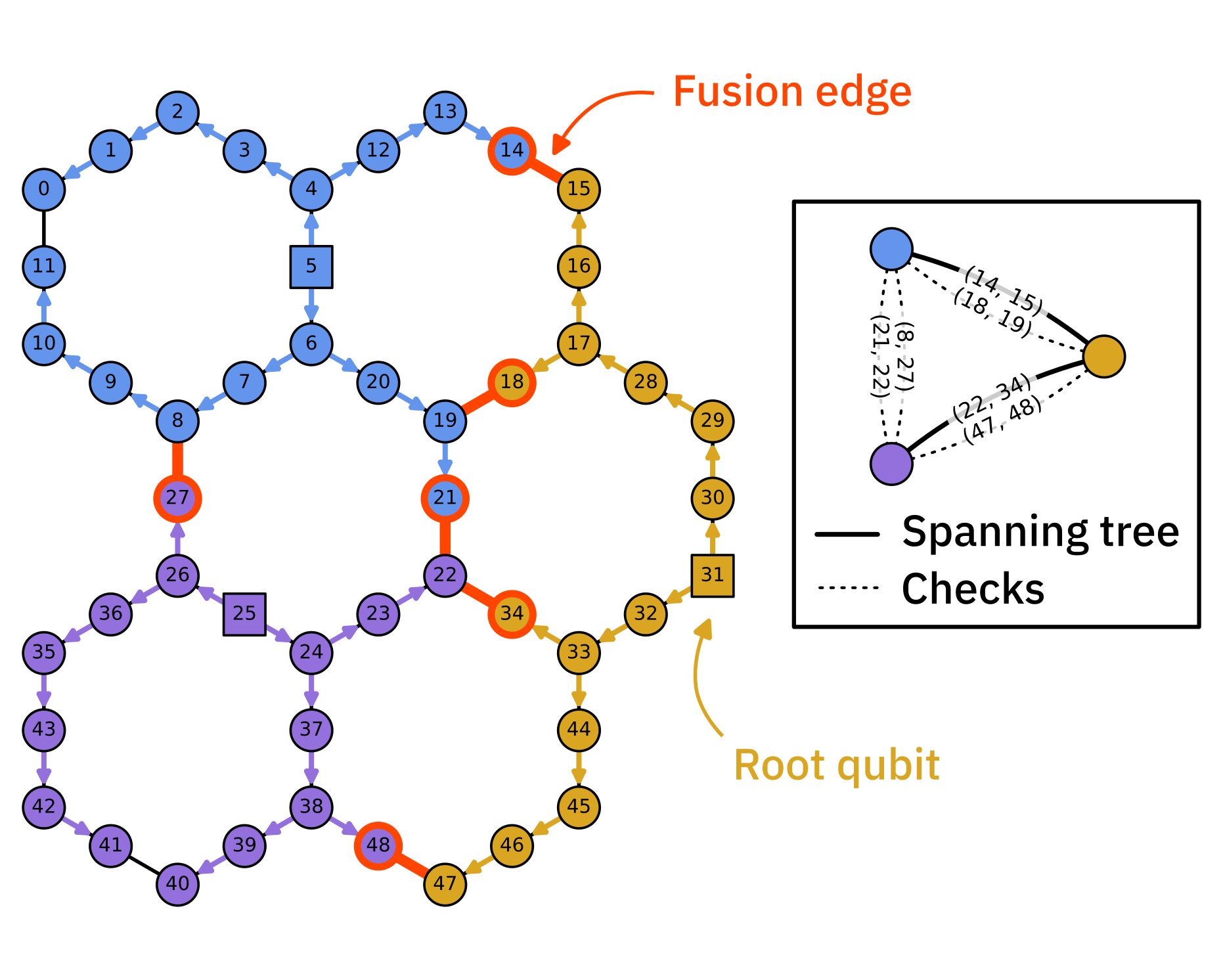}
    \caption{Setup for GHZ preparation on a heavy-hexagonal lattice with $k=3$ patches. Independently prepared patches are shown as distinct colors with $m$ fusion edges marked in red and the measured qubits circled. Square nodes indicate the root qubit for each patch, while directed arrows trace the \textsc{cnot} ladders along which each patch is grown. The inset shows the meta-graph, indicating the spanning tree of $k-1$ fusion edges (bold), together with the remaining $m-k+1$ edges (dashed) which serve as checks for error detection.}
    \label{fig:app_ghz}
\end{figure}

Contrast this with GHZ preparation on a 2D graph. As a concrete example, Fig.~\ref{fig:app_ghz} shows the setup on a heavy-hexagonal lattice with $k=3$. Each of the three patches is drawn in a distinct color, with fusion edges marked in red (and the measured fusion qubit circled). Within each patch, the GHZ state is grown along a minimal spanning tree. The root -- i.e., the qubit receiving the initial Hadamard -- is shown as a square, and \textsc{cnot}-ladder paths are indicated by arrows. 

The key distinction from the 1D case is best understood by considering the \emph{meta-graph} (inset of Fig.~\ref{fig:app_ghz}), in which each patch is contracted to a single node and the $m$ fusion edges become the edges between nodes. Merging the patches into a single GHZ state requires fusing along only a spanning tree of this meta-graph, which fixes both the minimal set of $k-1$ fusion measurements needed and the paths along which parities are accumulated for the conditional feedforward. Unlike in 1D, however, a 2D graph is generally endowed with additional fusion edges. In the present example, $m=6$, leaving $m-k+1 = 4$ edges outside the spanning tree. 

In the error-free case these extra edges carry information entirely redundant with the tree, since each closes a cycle whose total $ZZ$ parity is fixed. In the presence of noise, however, this redundancy presents an opportunity for error detection, as each surplus edge serves as a consistency check for postselection. Alternatively, as proposed in Ref.~\cite{Waring_RobustGHZ_2026}, at sufficiently high check density the redundant outcomes can be combined by majority vote, leading to more robust feedforward corrections.

\section{Error-detected preparation of Dicke states}
\label{app:dicke}

In this Appendix, we provide further details on the protocol employed for preparing Dicke states in Sec.~\ref{sssec:w_and_dicke}, which closely follows that of Ref.~\cite{Piroli_ApproximatingManyBody_2024}, here augmented with error detection. For the sake of clarity, we break the protocol into two distinct steps: (1) Preparation of the initial distribution, and (2) Projection onto a target excitation number $k$.

\subsection{Preparation of the initial distribution}
The first stage of the protocol is to prepare the product state,

\begin{equation}
\begin{split}
|\Psi_0\rangle &= \left(R_Y(\theta)\right)^{\otimes N}\ket{0}^{\otimes N} \\
&= \left[\cos \frac{\theta}{2}\ket{0} + \sin\frac{\theta}{2}\ket{1}\right]^{\otimes N} \\
&= \sum_{s=0}^{N}\sqrt{P(s)}\ket{D^N_s},
\end{split}
\end{equation}
where on the last line we have decomposed $\ket{\Psi_0}$ into a superposition of Dicke states weighted by the distribution
\begin{equation}
P(s) = \left(\cos^2 \frac{\theta}{2}\right)^{N-s}\left(\sin^2 \frac{\theta}{2}\right)^s\binom{N}{s}.
\end{equation}

To maximize the probability of successful projection onto the target state $|D^N_k\rangle$, we set $\sin^2(\theta/2) = k/N$, yielding the optimal rotation angle
\begin{equation}
\theta = 2\sin^{-1}\left(\sqrt{\frac{k}{N}}\right).
\end{equation}

We note that, for the special case $k=1$ corresponding to the W state, a smaller angle can be chosen (as in Sec.~\ref{ssec:wstate}) to increase fidelity at the expense of success probability.

\subsection{Projection onto target excitation number $k$}
Throughout this subsection (and this subsection only), we use hats to distinguish operators from scalars. Projection onto a particular excitation sector is implemented by estimating the phase
\begin{equation}
\hat{\varphi} = \frac{\hat{n} - k}{\tau},
\end{equation}
where $\hat{n} = \sum_{j=1}^N (1-\hat{Z}_j)/2$ is the total excitation number, $k$ the target excitation number, and $\tau$ a rescaling parameter. Because phase estimation resolves $\hat{\varphi}$ only modulo $1$, the excitation number $\hat{n}$ is correspondingly determined only modulo $\tau$, with excitation numbers differing by a multiple of $\tau$ aliased onto a common phase. While setting $\tau = N+1$ will eliminate aliasing entirely, the sharp concentration of the binomial weights about $k$ means that a much smaller $\tau$ renders the aliased contributions negligible. In practice it suffices to take $\tau$ to span a few multiples of the standard deviation $\sigma = \sqrt{k(1-k/N)}$. Since $\sigma\leq \sqrt{k}$ for all $k$, this choice gives $\tau = O(\sqrt{k})$ and hence $r = \log (\tau) = O(\log k)$ rounds of \textsc{distribute}--\textsc{collapse}, in contrast to the $O(\log N)$ rounds needed for exact preparation~\cite{Piroli_ApproximatingManyBody_2024}.

While standard quantum phase estimation (QPE) suffices to estimate $\hat{\varphi}$, here we highlight an alternative approach reminiscent of iterative phase estimation~\cite{Kitaev_QuantumMeasurements_1995, Dobsicek_ArbitraryAccuracy_2007} that eliminates the inverse quantum Fourier transform of standard QPE, reduces the total ancilla count, and is better suited to hardware connectivity constraints as it enables reuse of a single GHZ register of size $O(N)$ across rounds rather than preparing and using $O(\log k)$ such registers in parallel. Moreover, unlike standard iterative phase estimation, no feedforward is needed between rounds as our goal is to project onto a predetermined excitation number $k$. 

To explain, it is helpful to decompose the phase operator into binary,
$\hat{\varphi} = 0.\hat{b}_1 \hat{b}_2 \hat{b}_3 \ldots \hat{b}_r$, where $\tau = 2^r$ and unhatted $b_j\in\{0,1\}$ denote measurement outcomes. We define the unitary $U = \exp(2\pi i \hat{\varphi})$. The projection protocol proceeds by iteratively estimating $\hat{\varphi}$ bit-by-bit, beginning with the least-significant bit $b_r$ and ending with the most significant $b_1$. Because $\hat{\varphi}$ is shifted relative to the desired excitation number $k$, successful projection entails measuring $b_j = 0$ for all $j\in\{1,2,\ldots r\}$. Consequently, the phase corrections of standard iterative phase estimation are unnecessary: either the projection succeeds and the corrective phase is $e^{-i\pi \cdot 0} = 1$, or it fails and the shot is discarded. Said another way, under the condition that $b_i = 0$ for $i = j+1, j+2, \ldots r$, bit $b_j$ can be extracted via a simple Hadamard test:
\begin{equation}
\includegraphics[width = 0.95\linewidth]{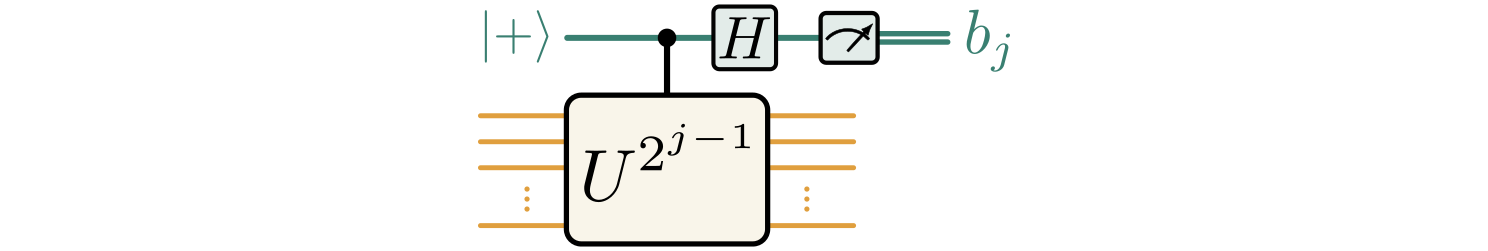}
\label{eq:bj_extract}
\end{equation}
As shown in Fig.~\ref{fig:gates}(d), our \textsc{distribute}--\textsc{collapse} framework enables an error-detected, constant-depth implementation of this subcircuit, requiring $O(N)$ ancillas and a single round of feedforward. 

As established in Ref.~\cite{Piroli_ApproximatingManyBody_2024}, the subcircuits for extracting all $r$ bits can be combined in several ways with distinct spacetime tradeoffs. A sequential implementation reuses a single GHZ register across rounds,  requiring $O(N)$ ancillas and $O(\log k)$ rounds of \textsc{distribute}--\textsc{collapse}, each of constant depth. Alternatively, parallelizing all $r$ Hadamard tests yields an $O(1)$-depth approach at the cost of $O(N\log k)$ ancillas. While either can be augmented with our error detection scheme, we expect the former to be best suited to near-term devices with connectivity constraints and limited qubit counts.

\section{Simulation details}
\label{app:noise_model}

\begin{figure*}
\begin{tikzpicture}[
  font=\small\color{ink},
  % main pipeline stages
  mainbox/.style={
    rectangle, rounded corners=5pt,
    draw=teal, line width=1pt, fill=teallt,
    text width=6.6cm, align=center,
    minimum height=1.05cm, inner sep=8pt
  },
  % input / output
  iobox/.style={
    rectangle, rounded corners=5pt,
    draw=teal, line width=1pt, fill=teal,
    text width=6.6cm, align=center,
    minimum height=0.85cm, inner sep=7pt,
    text=white, font=\small\bfseries
  },
  % timing / parameter side boxes (gold/cream)
  parambox/.style={
    rectangle, rounded corners=4pt,
    draw=gold, line width=0.9pt, fill=cream,
    text width=3.35cm, align=left,
    inner sep=7pt, font=\scriptsize\color{ink}
  },
  % noise side box (red family)
  noisebox/.style={
    rectangle, rounded corners=4pt,
    draw=red, line width=0.9pt, fill=redlt,
    text width=3.7cm, align=left,
    inner sep=7pt, font=\scriptsize\color{ink}
  },
  mainarrow/.style={-{Stealth[length=6pt,width=6pt]}, line width=1.4pt, teal},
  sidearrow/.style={-{Stealth[length=5pt]}, line width=0.9pt, gold!80!black, densely dashed},
  sidearrowR/.style={-{Stealth[length=5pt]}, line width=0.9pt, red!75, densely dashed},
]

%──────── main pipeline (centre column) ────────
\node[iobox] (input) {Input: quantum circuit $\mathcal{C}$};

\node[mainbox, below=0.5cm of input] (transpile) {
  \hdr{Transpile to basis gates}\\[3pt]
  {\scriptsize\ttfamily \{id, x, y, z, h, s, sdg, cx, cz, swap, measure, reset\}}
};

\node[mainbox, below=0.5cm of transpile] (schedule) {
  \hdr{Schedule $+$ PadDelay}\\[3pt]
  {ALAP scheduling inserts \texttt{delay} into all idle qubit periods}
};

\node[mainbox, below=0.5cm of schedule] (convert) {
  \hdr{Convert delays to Pauli error channels}\\[5pt]
  $p_X = p_Y = \tfrac{1}{4}\!\left(1 - e^{-t/T_1}\right)$\\[3pt]
  $p_Z = \tfrac{1}{2}\!\left(1 - e^{-t/T_2}\right) - \tfrac{1}{4}\!\left(1 - e^{-t/T_1}\right)$
};

\node[mainbox, below=0.5cm of convert] (assemble) {
  \hdr{Assemble noise model $\mathcal{N}$}\\[3pt]
  {idle $+$ gate $+$ readout $+$ reset errors}
};

\node[iobox, below=0.5cm of assemble] (output)
  {Output: $\bigl(\mathcal{C}_{\mathrm{sched}},\;\mathcal{N}\bigr)$};

\draw[mainarrow] (input)     -- (transpile);
\draw[mainarrow] (transpile) -- (schedule);
\draw[mainarrow] (schedule)  -- (convert);
\draw[mainarrow] (convert)   -- (assemble);
\draw[mainarrow] (assemble)  -- (output);

%──────── right: gate durations → schedule ────────
\node[parambox, right=1.4cm of schedule] (params) {
  \hdrG{DURATIONS}\\[5pt]
  1Q gate\hfill $30\,\mathrm{ns}$\\[1pt]
  2Q gate\hfill $70\,\mathrm{ns}$\\[1pt]
  Measure\hfill $1\,\mu\mathrm{s}$\\[1pt]
  Reset\hfill $30\,\mathrm{ns}$\\[1pt]
  Feedforward\hfill $650\,\mathrm{ns}$
};
\draw[sidearrow] (params.west) -- (schedule.east);

%──────── left: coherence times → convert ────────
\node[parambox, left=1.4cm of convert] (coherence) {
  \hdrG{COHERENCE TIMES}\\[5pt]
  $T_1$: amplitude damping\\[2pt]
  $T_2$: dephasing
};
\draw[sidearrow] (coherence.east) -- (convert.west);

%──────── right: noise sources → assemble ────────
\node[noisebox, right=1.4cm of assemble] (noisecomp) {
  \hdrR{NOISE SOURCES}\\[5pt]
  Idle:\hfill $\mathcal{E}_{\mathrm{idle}}(t,T_1,T_2)$\\[2pt]
  1Q gates:\hfill depol.$(\varepsilon_{1\mathrm{Q}})$\\[2pt]
  2Q gates:\hfill depol.$(\varepsilon_{2\mathrm{Q}})$\\[2pt]
  Readout:\hfill $p(0|1),\,p(1|0)$\\[2pt]
  Reset:\hfill $X$-flip$(p_r)$
};
\draw[sidearrowR] (noisecomp.west) -- (assemble.east);

\end{tikzpicture}
\caption{Pipeline for the noisy simulations in this work.}
\label{fig:simulator}
\end{figure*}
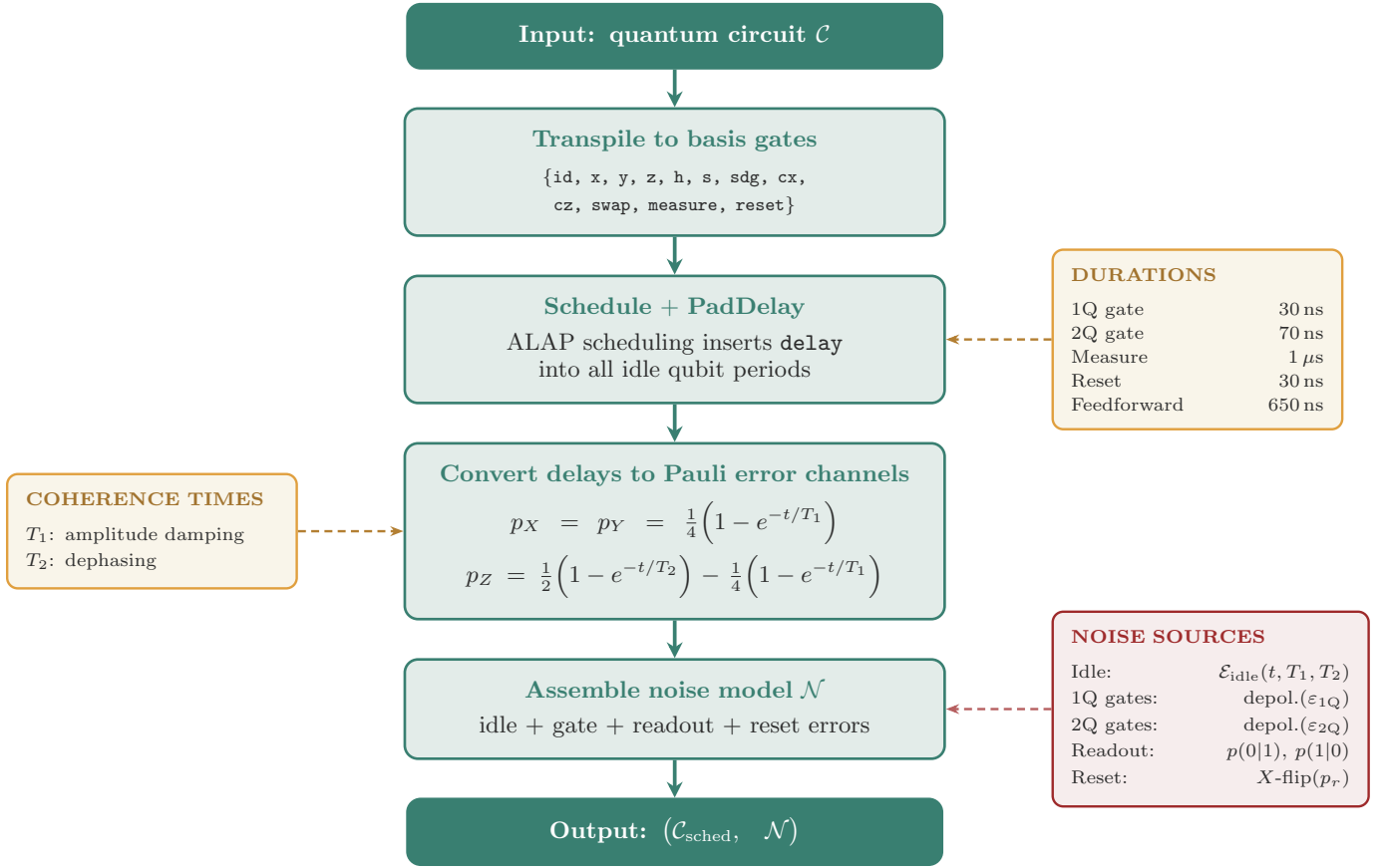

Fig.~\ref{fig:simulator} shows an outline of the simulation infrastructure used for the noisy simulations presented in Sec.~\ref{ssec:simulation}. To account for idling errors incurred during mid-circuit measurements, the transpiled circuit is scheduled using the as-late-as-possible (ALAP) policy. After scheduling, all idle periods in the circuit are identified and annotated with delays (\texttt{PadDelay}). 
Doing so allows the delays to be converted into error channels: using the Pauli-twirling approximation~\cite{Silva_ScalableProtocol_2008}, every delay is converted into a Pauli error channel with probabilities assigned according to delay duration.
Once the circuit is annotated with delays and the corresponding error channels, we build a custom noise model that incorporates gate, measurement, idling, and reset errors, and carry out simulations using the matrix product state backend in Qiskit. Model parameters are reported in Table~\ref{tab:device_params}, chosen to loosely approximate error rates on \texttt{ibm\_boston}.

\begin{table}[h]
\centering
\begin{tabular}{llr}
\hline
\textbf{Parameter} & \textbf{Symbol} & \textbf{Value} \\
\hline
Relaxation time            & $T_1$              & $250~\mu\mathrm{s}$        \\
Coherence time             & $T_2$              & $250~\mu\mathrm{s}$        \\
Single-qubit gate duration & $\tau_{1\mathrm{qb}}$   & $30~\mathrm{ns}$      \\
\textsc{cz} gate duration    & $\tau_{\textsc{cz}}$   & $70~\mathrm{ns}$      \\
Single-qubit gate error    & $\epsilon_{1\mathrm{qb}}$    & $1\times10^{-4}$ \\
Two-qubit gate error       & $\epsilon_{2\mathrm{qb}}$    & $1.5\times10^{-3}$ \\
\hline
\end{tabular}
\caption{Model parameters used for the noisy simulations in Sec.~\ref{ssec:simulation}.}
\label{tab:device_params}
\end{table}

\begin{figure}
    \centering
    \includegraphics[width=\linewidth]{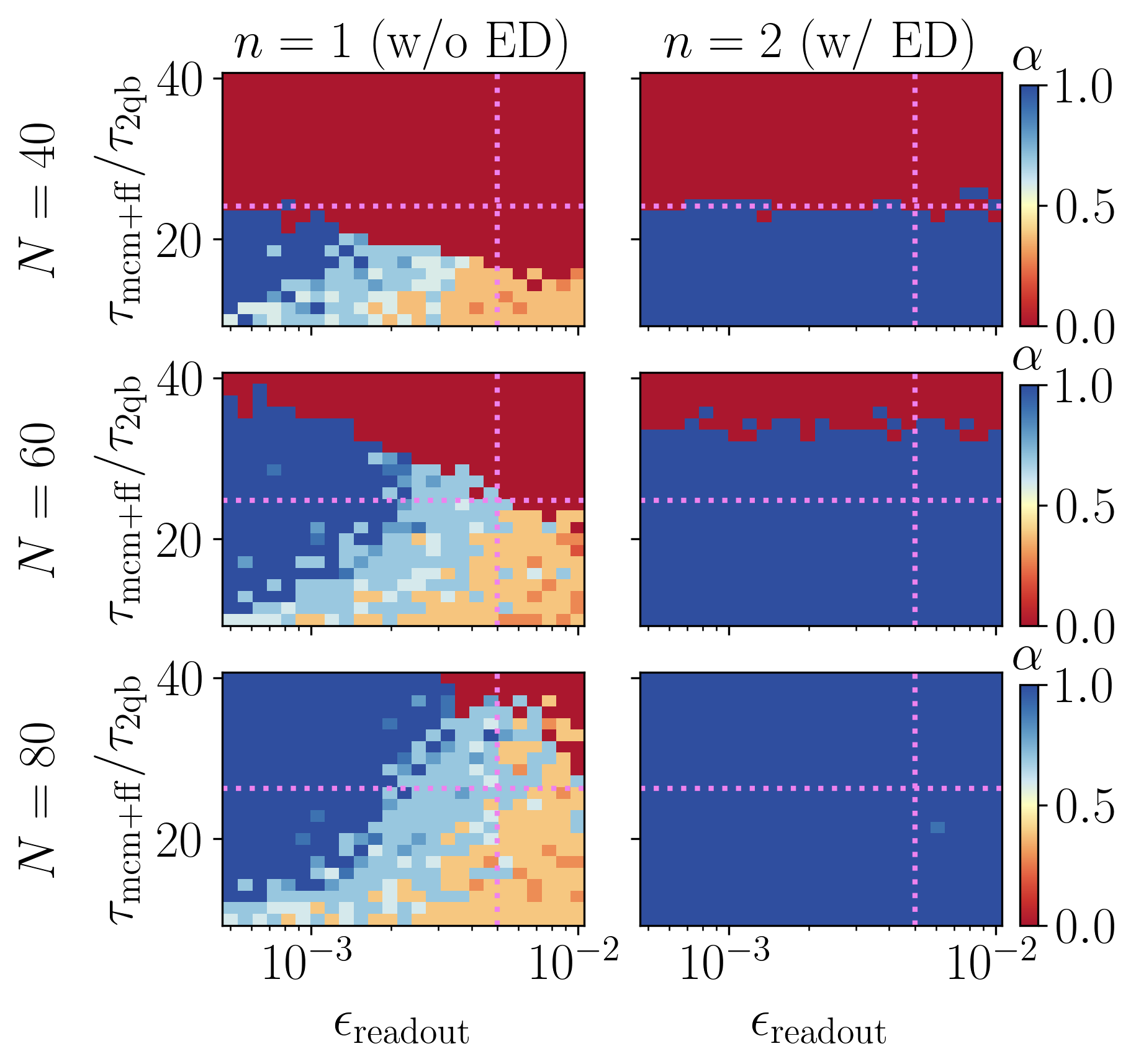}
    \caption{The measurement usage ratio $\alpha = m_{\mathrm{opt}}/(\frac{N}{2} - 1)$, indicating the optimal strategy for preparing the GHZ state with ($n=2$, implicit checks) and without ($n=1$) error detection for three system sizes: $N = 40$ (top row), $N = 60$ (middle row), $N = 80$ (bottom row). Dotted pink lines indicate the ($N$-dependent) mid-circuit measurement and feedforward duration and readout error on \texttt{ibm\_boston} estimated via scheduling.
    See Fig.~\ref{fig:simulation} and surrounding text for details.}
    \label{fig:alpha_stack}
\end{figure}

As an accompaniment to Fig.~\ref{fig:simulation} of the main text, Fig.~\ref{fig:alpha_stack} shows the optimal GHZ preparation strategy as a function of $\tau_{\mathrm{mcm+ff}}/\tau_{\mathrm{2qb}}$ $\epsilon_{\textrm{readout}}$ for $N=40,\, 60,\, $ and $80$. For both this figure and Fig.~\ref{fig:simulation} of the main text, horizontal dashed pink lines -- indicating the expected duration of mid-circuit measurement and feedforward normalized to two-qubit gate duration -- were extracted from the circuit schedule after transpilation on \texttt{ibm\_boston}. For the latter quantity, we use the duration associated with adding one additional qubit to a GHZ state: $\tau_{\mathrm{2qb}} = \tau_{\textsc{cz}} + \tau_{\textrm{Hadamard}} = 100$ ns. Note that although a \textsc{cnot} gate decomposes into a \textsc{cz} gate and two Hadamard gates, successive layers can be partially parallelized such that only one counts against the total duration.

\section{Experimental Details}
\label{app:lrcx}
\label{app:wstate}

\begin{figure*}[htbp]
    \centering
    \includegraphics[width=\textwidth]{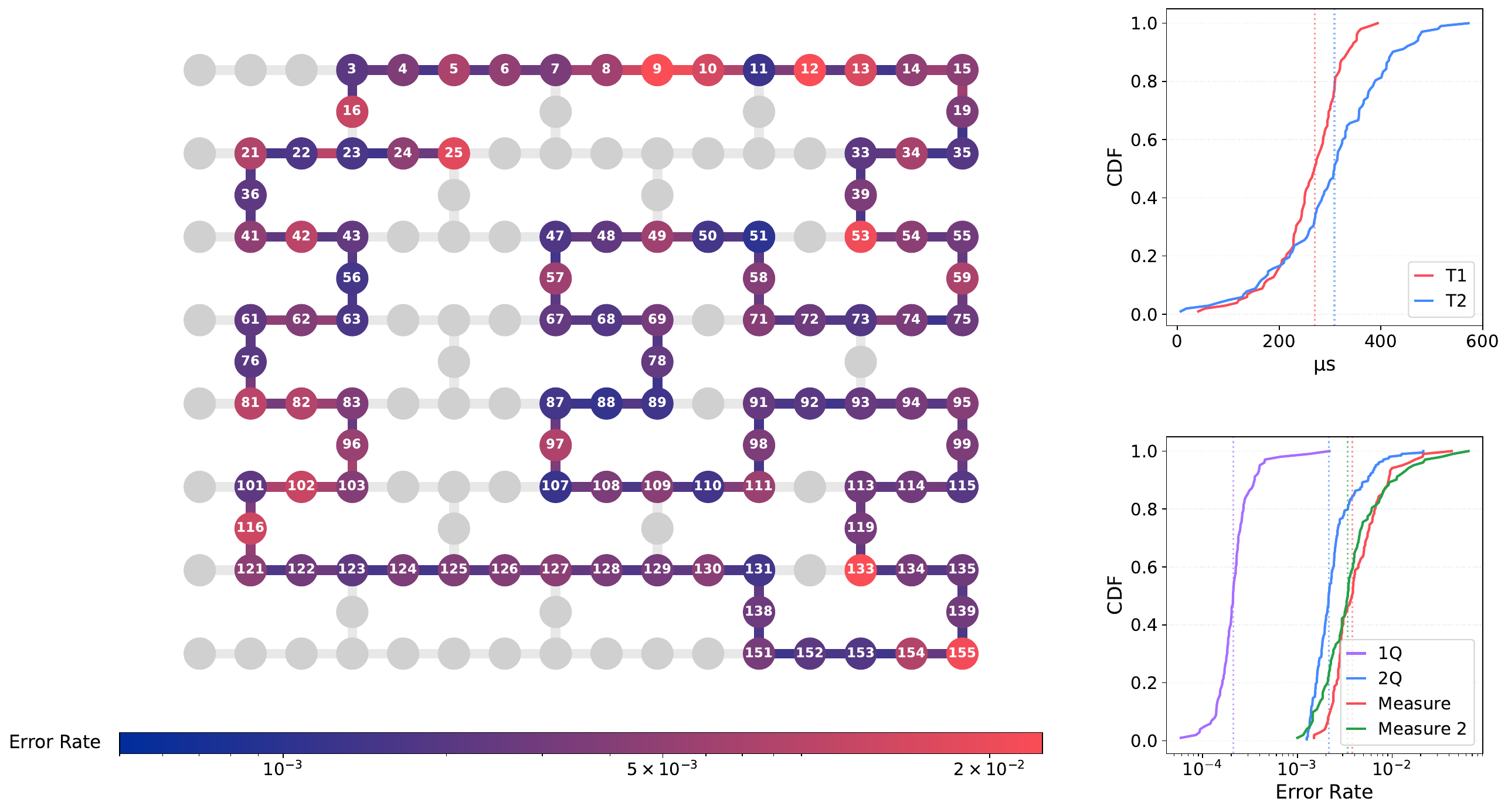}
    \caption{Device details for the long-range \textsc{cnot} distance-sweep experiment [Fig. \ref{fig:experiment} (a)]. Left: qubit layout on the Heron r3 heavy-hex topology showing the 102-qubit chain used for distances $\ell \in \{4, 20, 36, 52, 68, 84, 100\}$. Node color indicates \texttt{measure\_2} readout error; edge color indicates two-qubit gate error from layer fidelity estimation (LFE), both on a shared logarithmic scale. Top right: cumulative distribution of $T_1$ and $T_2$ coherence times for the selected qubits. Bottom right: cumulative distribution of single-qubit (SX), two-qubit (LFE), standard measurement, and \texttt{measure\_2} error rates.}
    \label{fig:lrcx_n_sweep_layout}
\end{figure*}

\begin{figure*}[htbp]
    \centering
    \includegraphics[width=\textwidth]{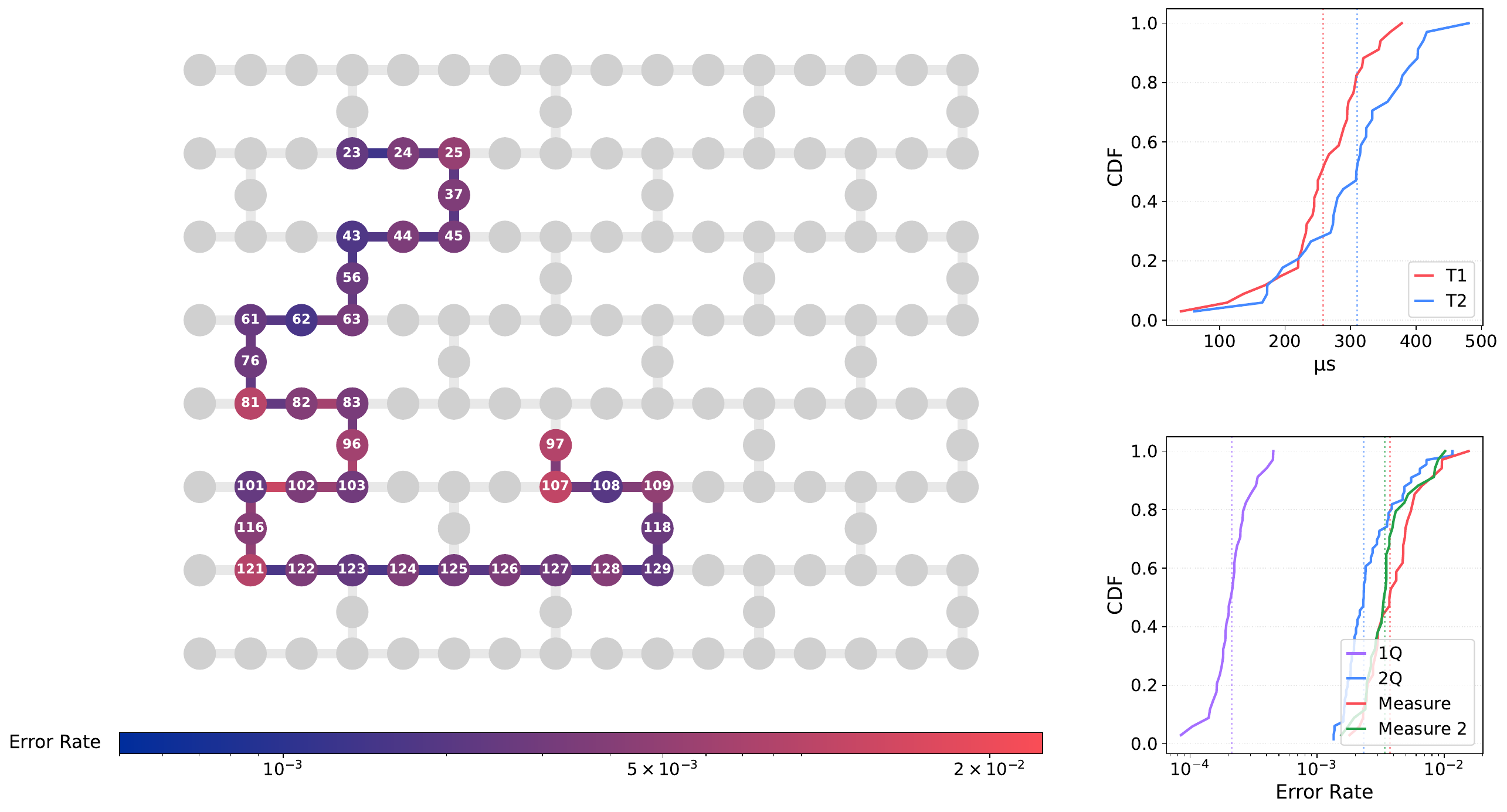}
    \caption{Device details for the long-range \textsc{cnot} code-size sweep [Fig. \ref{fig:experiment} (b)]. The 34-qubit subchain (32 intermediary + 2 system qubits) is used for variable error-detection code sizes on a fixed physical layout. Node color indicates \texttt{measure\_2} readout error.}
    \label{fig:lrcx_d_sweep_layout}
\end{figure*}

\begin{figure*}[htbp]
    \centering
    \includegraphics[width=\textwidth]{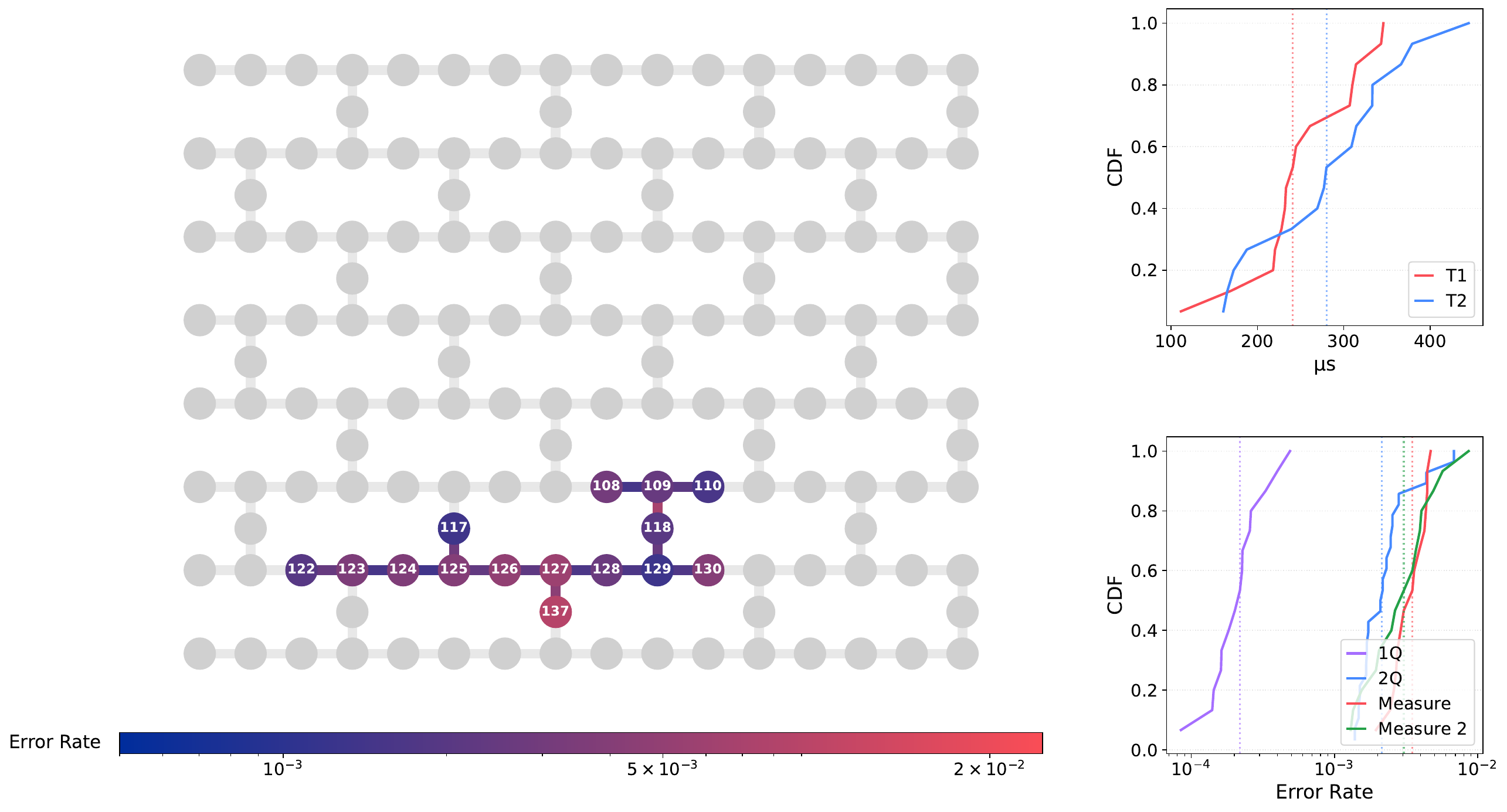}
    \caption{Device details for the $N=5$ W state DFE experiment [Fig. \ref{fig:experiment} (c) and (d)]. The 15-qubit region (10-qubit GHZ chain + 5 data qubits) corresponds to layout shift 2. Node color indicates \texttt{measure\_2} readout error; edge color indicates two-qubit gate error from LFE.}
    \label{fig:wstate_N5_layout}
\end{figure*}

\begin{figure*}[htbp]
    \centering
    \includegraphics[width=\textwidth]{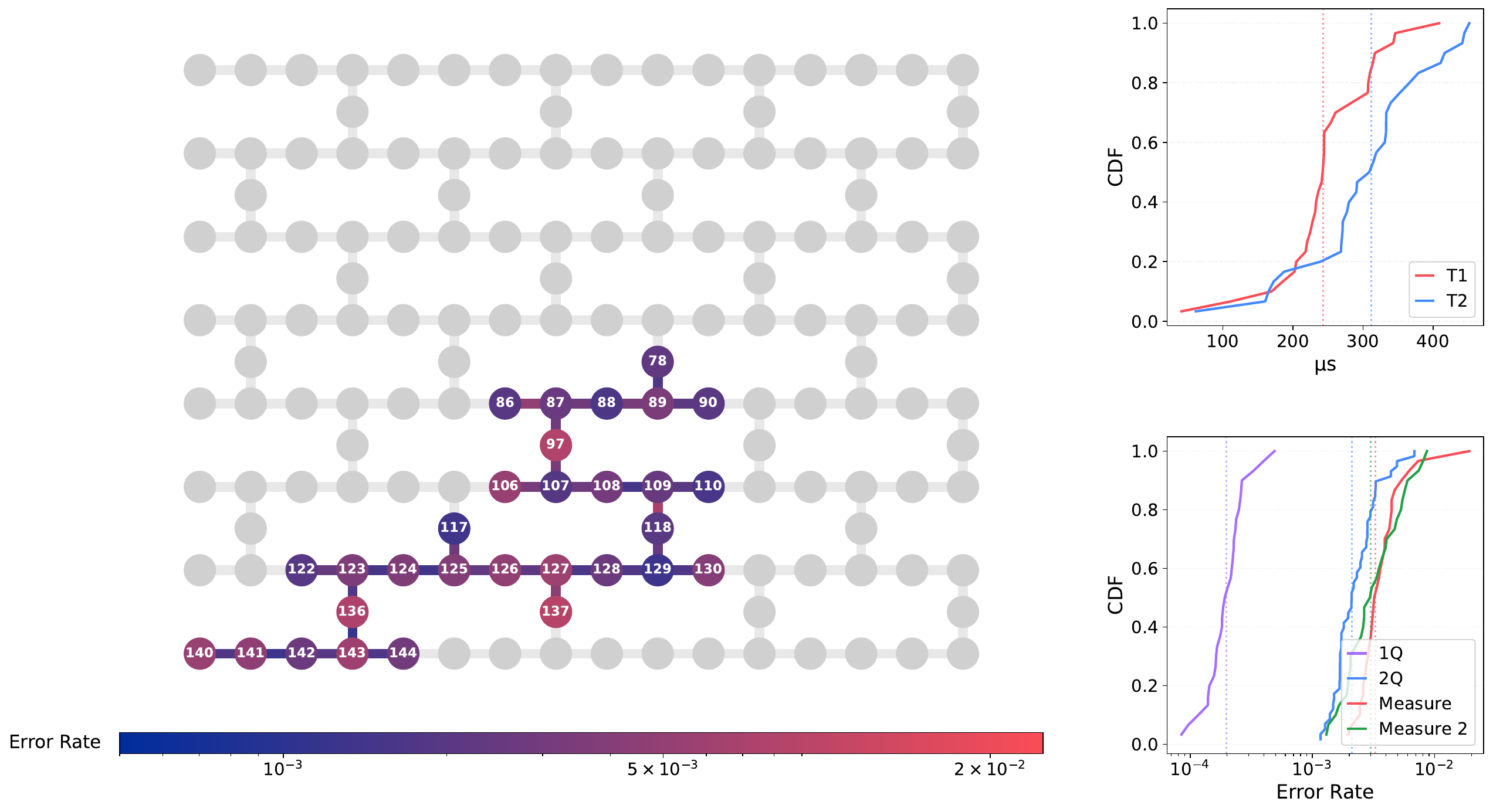}
    \caption{Device details for the $N=10$ W state DFE experiment. The 30-qubit region (20-qubit GHZ chain + 10 data qubits) corresponds to layout shift 0.}
    \label{fig:wstate_N10_layout}
\end{figure*}

\begin{figure*}[htbp]
    \centering
    \includegraphics[width=\textwidth]{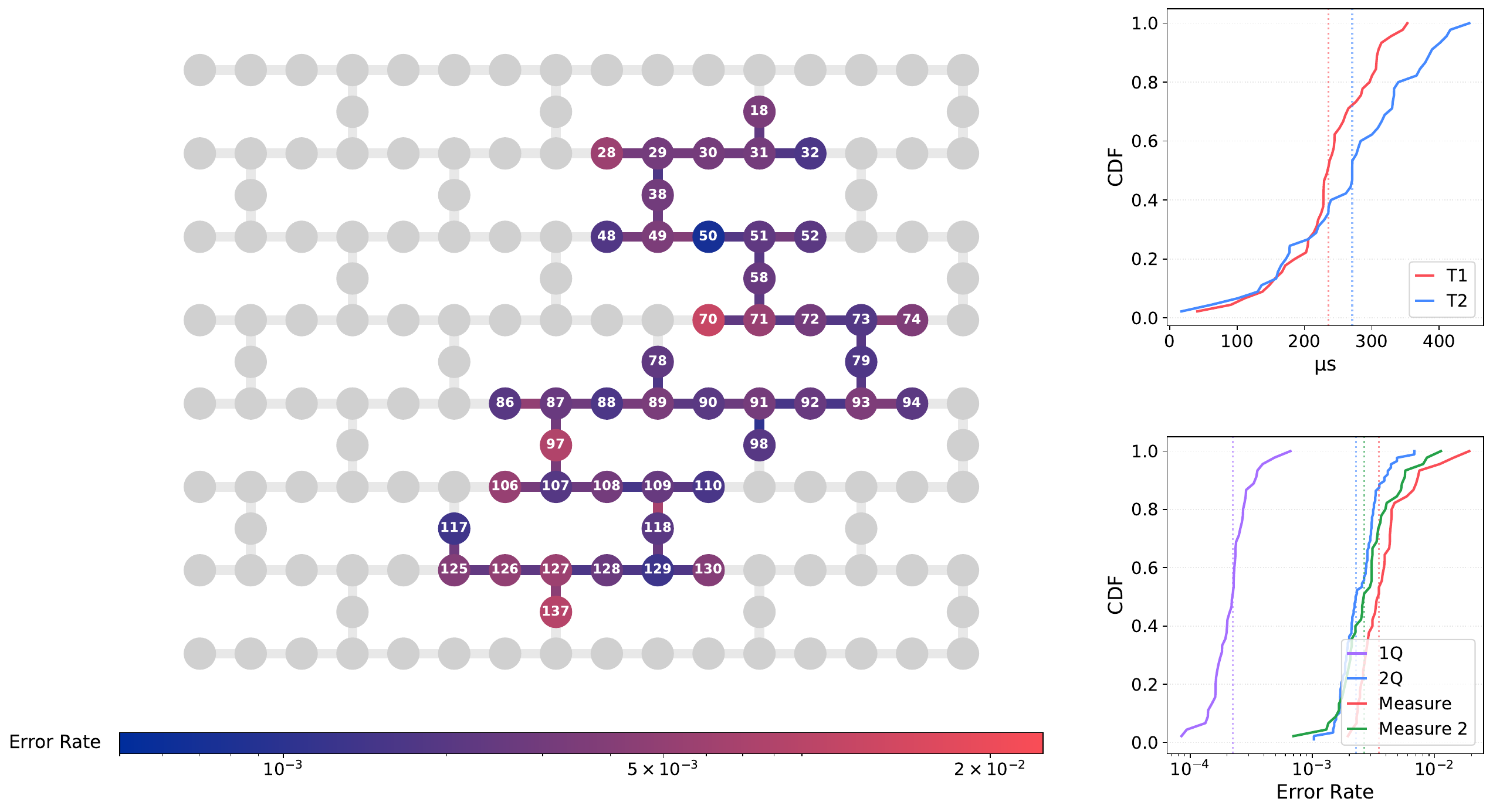}
    \caption{Device details for the $N=15$ W state DFE experiment. The 45-qubit region (30-qubit GHZ chain + 15 data qubits) corresponds to layout shift 3.}
    \label{fig:wstate_N15_layout}
\end{figure*}

\begin{figure*}[htbp]
    \centering
    \includegraphics[width=\textwidth]{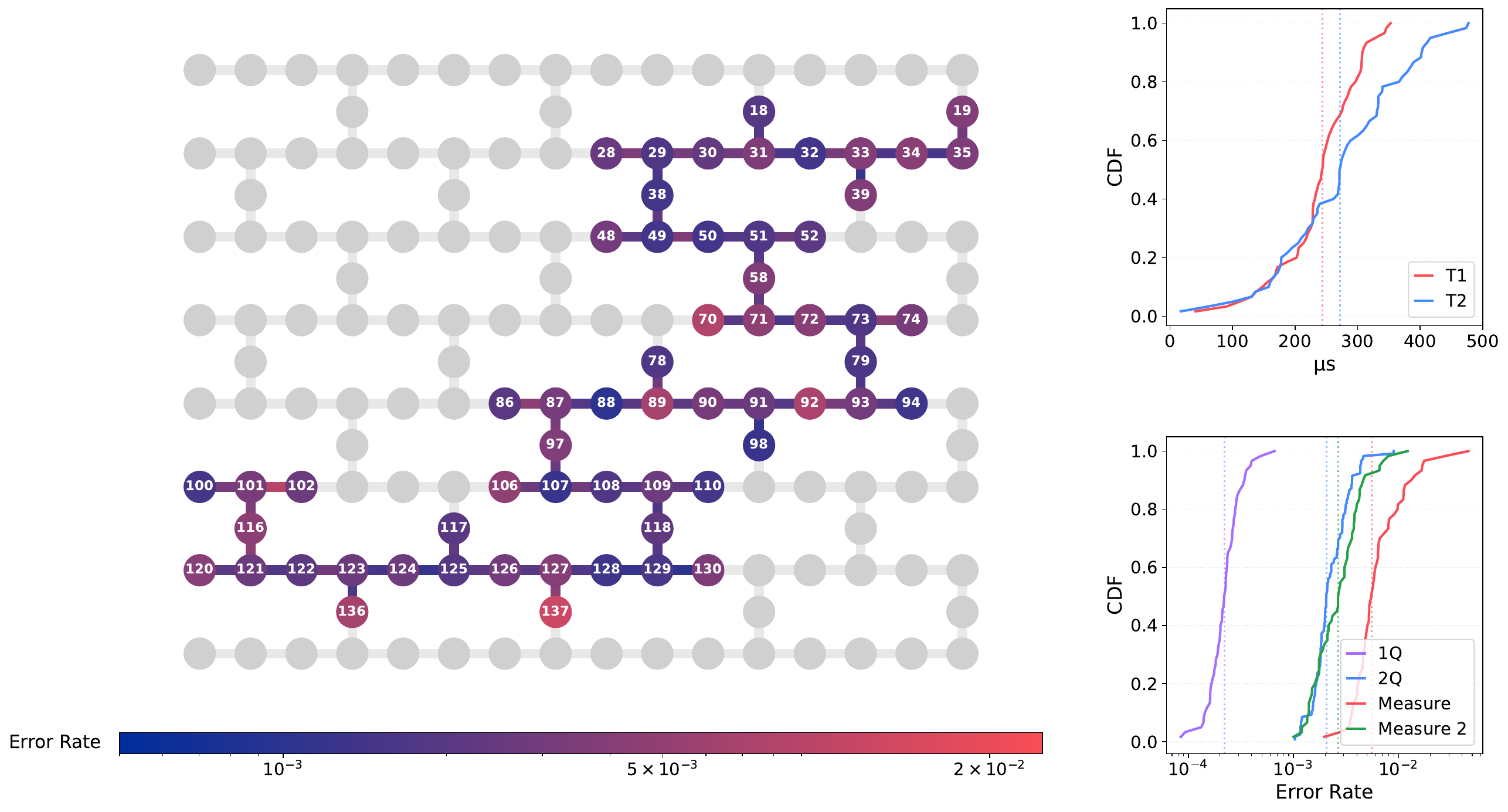}
    \caption{Device details for the $N=20$ W state DFE experiment. The 60-qubit region (40-qubit GHZ chain + 20 data qubits) uses a different physical chain than $N \leq 15$, selected from a separate calibration run.}
    \label{fig:wstate_N20_layout}
\end{figure*}

Quantum experiments for the long-range \textsc{cnot} [Sec.~\ref{ssec:lrcx}] and W state preparation [Sec.~\ref{ssec:wstate}] were run on \texttt{ibm\_boston}, a Heron r3 superconducting quantum processor with 156 qubits arranged in a heavy-hex lattice. Qubit selection was done using mapomatic \cite{PRXQuantum.4.010327}, by using a proxy fidelity $F_{\text{objective}} = \prod_{g_1, g_2, m} (1-e_{g_1})  (1-e_{g_2}) (1-e_m) $ as the objective function. Here single-qubit gate errors $e_{g_1}$ were estimated using randomized benchmarking (RB) \cite{magesan2011scalable}, two-qubit gate errors $e_{g_2}$ were computed using layer fidelity RB \cite{mckay2023benchmarking}, and measurement errors $e_m$ were computed by capturing the confusion matrix for each qubit.  The qubit layouts for each experiment, along with the respective two-qubit gate errors, measurement errors, relaxation times and coherence times, are shown in Figs.~\ref{fig:lrcx_n_sweep_layout}--\ref{fig:lrcx_d_sweep_layout} and Figs.~\ref{fig:wstate_N5_layout}--\ref{fig:wstate_N20_layout}. 

Initial state preparation was enhanced via preselection, a protocol that initializes qubits in the ground state, measures them immediately, and discards shots where any qubit deviates from the expected ground state. \verb|measure_2|, which has a measurement pulse shorter than  \verb|measure|, was used for mid-circuit measurement and \verb|measure| for terminal measurements; other than for $N=20$ W state experiments, where we utilized \verb|measure_2| to perform the terminal measurement as well.

Dynamical decoupling~\cite{viola1998dynamical,viola1999dynamical,duan1999suppressing,zanardi1999symmetrizing,lidar2014review} is known to be effective on superconducting qubit platforms ~\cite{pokharel2018demonstration,ezzell2023dynamical}. Here for idle gaps that occur in parallel with unitary operations, we apply a $XX$ dynamical decoupling sequence. A crucial exception is made for idle gaps that occur during measurement layers -- two $XX$ pulses are applied, first covering the duration of the measurement pulse and the next covering the duration between the measurement pulse and the next non-idle operation. \verb|stretch| functionality is used to make the second $XX$ pulse agnostic to the time taken to perform any classical feedforward operations. This sequence is similar to but simpler than the sequence used in Ref.~\cite{baumer2024quantum}, where an XY-4 sequence was used instead of the simpler $XX$ pulse. Optimization of error suppression for dynamic circuits, using either more sophisticated deterministic sequences or by tailoring the sequences \cite{tong2026learning} could further bolster performance of the primitives introduced here.

\begin{figure*}[htbp]
    \centering
    \includegraphics[width=\textwidth]{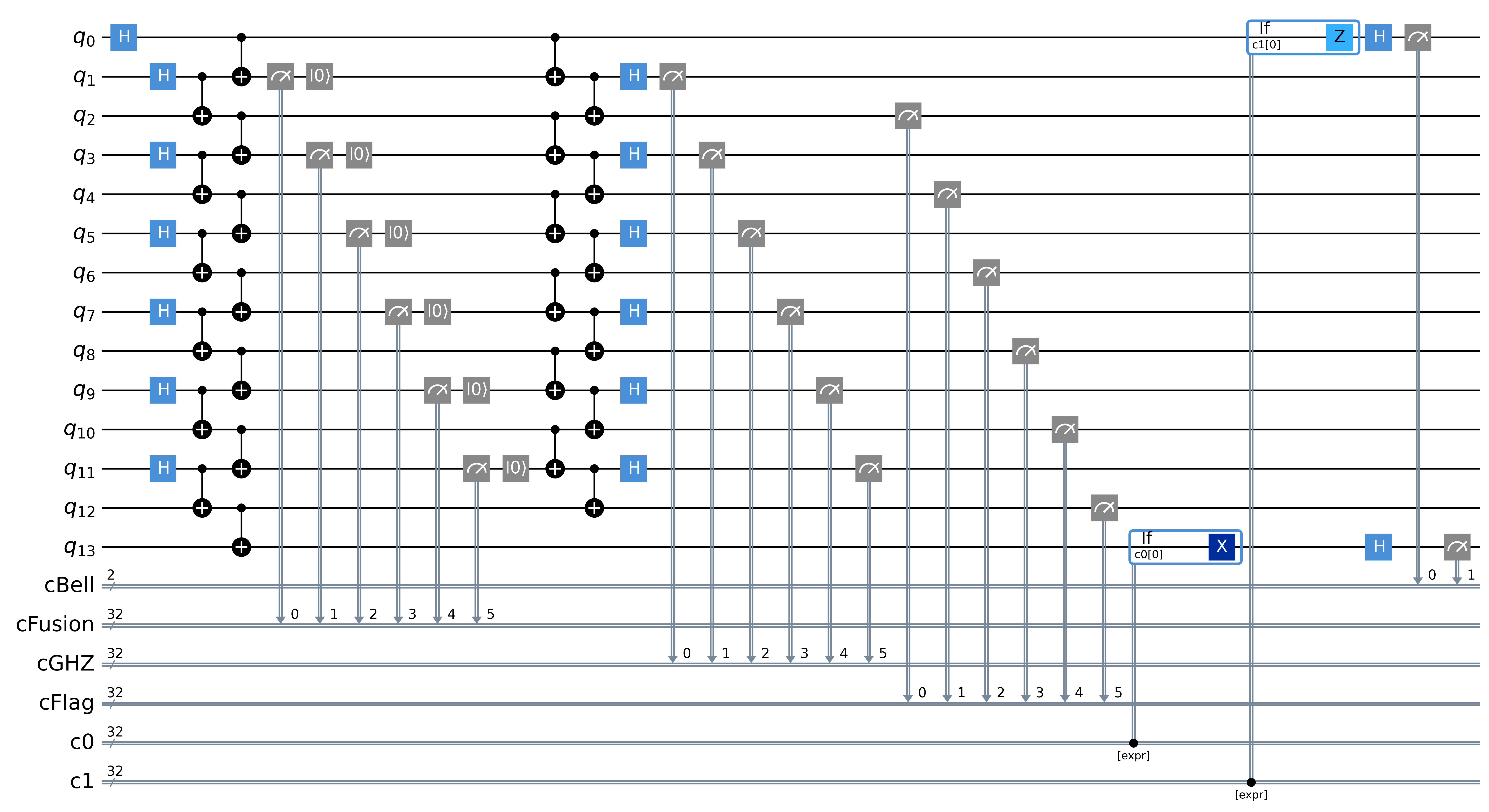}
    \caption{Long-range \textsc{cnot} circuit with explicit checks at distance $\ell=12$. The structure is identical to Fig.~\ref{fig:lrcx_circuit_rep}, but error detection uses a single \textsc{cnot}--\textsc{h} sequence per flag qubit, corresponding to an explicit check.}
    \label{fig:lrcx_circuit_unc}
\end{figure*}

\begin{figure*}[htbp]
    \centering
    \includegraphics[width=\textwidth]{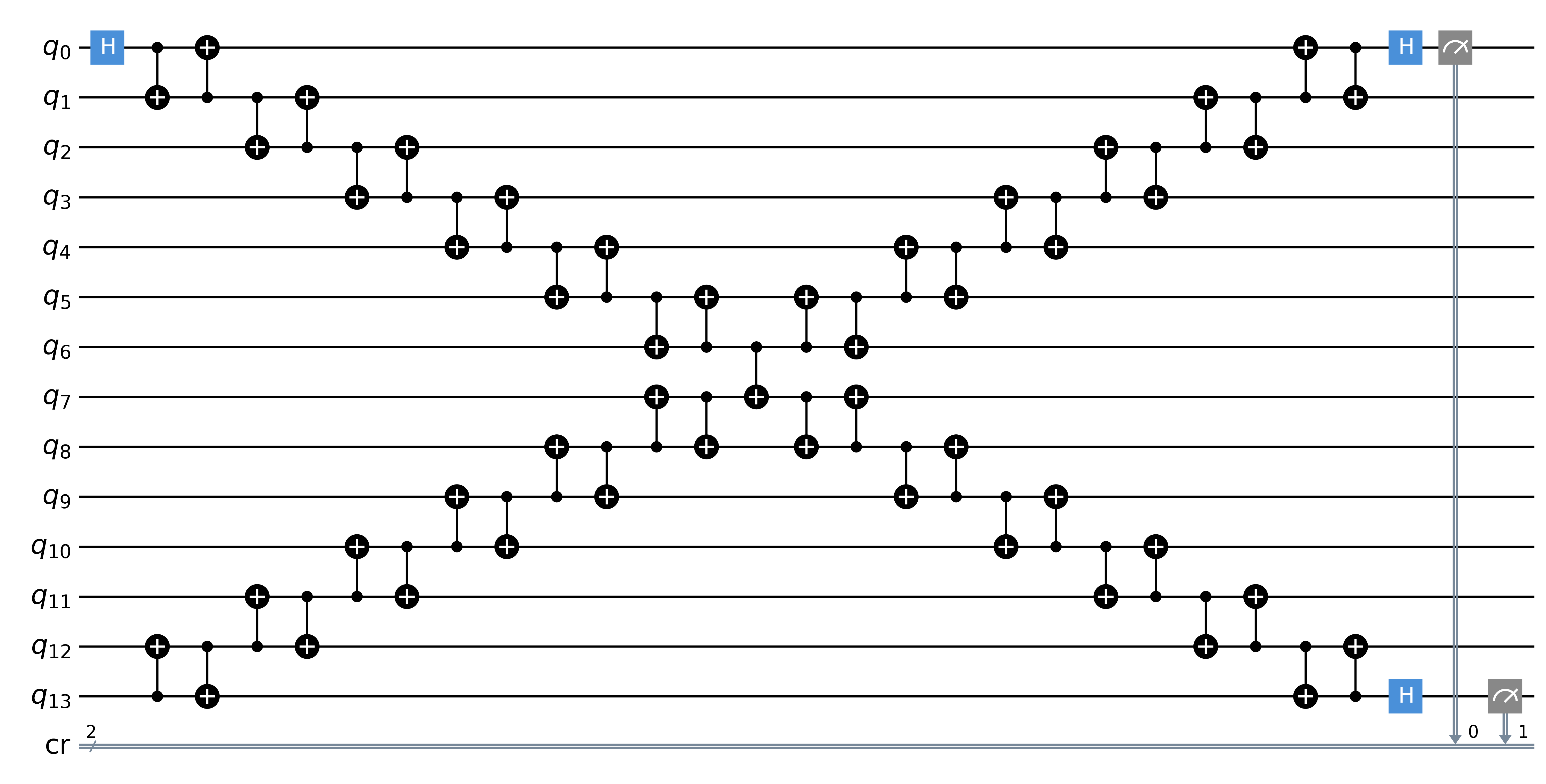}
    \caption{Long-range \textsc{cnot} circuit for distance $\ell=12$ using local unitary gates. The control and target qubit's states propagate inward via a \textsc{swap} ladder, which can be implemented using two \textsc{cnot} gates per \textsc{swap} as the intermediary qubits are in the ground state. A single \textsc{cnot} is applied at the meeting point before the \textsc{swap} ladder is carried out in reverse. The two-qubit gate depth scales linearly with $\ell$.}
    \label{fig:lrcx_circuit_unitary}
\end{figure*}

\begin{figure*}[htbp]
    \centering
    \includegraphics[width=\textwidth]{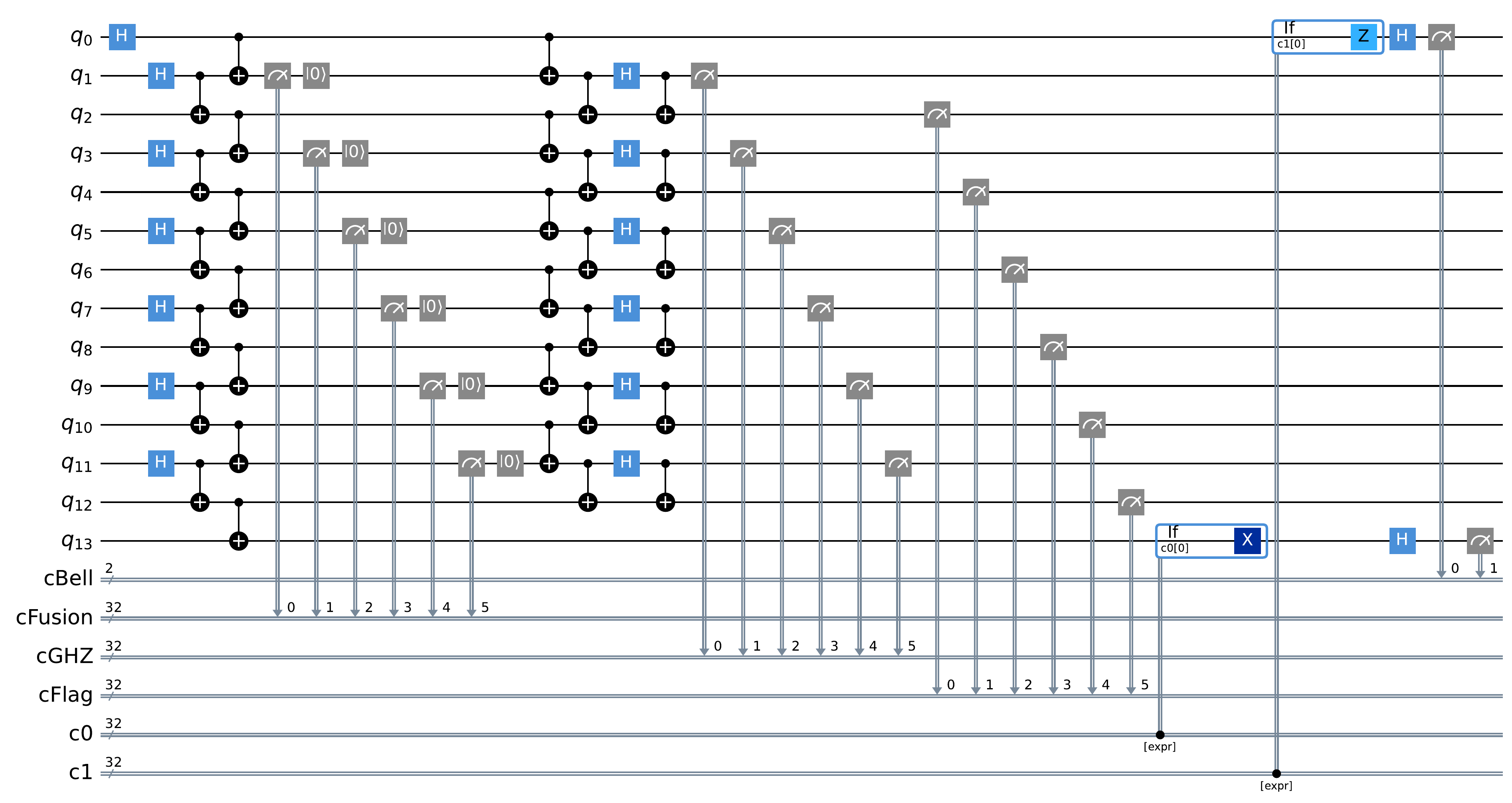}
    \caption{Long-range \textsc{cnot} circuit with implicit checks at distance $\ell=12$ (14 qubits total), measured in the XX basis. Bell pairs are prepared across intermediary qubits, fusion measurements are performed with mid-circuit measurement and reset, and $n=2$ implicit checks (\textsc{cnot}--\textsc{h}--\textsc{cnot}) are carried out before the final Bell measurement.}
    \label{fig:lrcx_circuit_rep}
\end{figure*}

Long-range \textsc{cnot} experiments [Sec.~\ref{ssec:lrcx}, Fig.~\ref{fig:experiment}(a)] were performed on a linear chain of qubits and used to prepare Bell states $|\Phi^+\rangle$ at graph distances $\ell \in \{4, 20, 36, 52, 68, 84, 100\}$ with total physical qubit counts of $\ell + 2$ per circuit. Representative circuits for $\ell = 12$ with $n=2$ implicit and explicit checks are shown in Figs.~\ref{fig:lrcx_circuit_rep} and \ref{fig:lrcx_circuit_unc}, respectively. The corresponding static unitary implementation is shown in Fig. \ref{fig:lrcx_circuit_unitary}. Each circuit was measured in three Pauli bases ($XX$, $YY$, $ZZ$) to determine the Bell fidelity $F_\text{Bell} = \frac{1}{4}(1 +\langle XX \rangle - \langle YY \rangle + \langle ZZ \rangle) $. Shot counts were scaled with problem size: $3 \times 10^4$ shots for $\ell \leq 20$, $6 \times 10^4$ for $\ell = 36$, $1.5 \times 10^5$ for $\ell \in \{52, 68\}$, and $3 \times 10^5$ for $\ell \in \{84, 100\}$.  For the $n$-sweep experiments [Fig.~\ref{fig:experiment} (b)], we fixed the long-range \textsc{cnot} distance at $\ell = 32$ and varied the code size $n$ for both strategies of explicit and implicit checks. Here each experiment was performed with $10^5$ shots per circuit. Device characterization for each experiment is provided in Figs.~\ref{fig:lrcx_n_sweep_layout} and \ref{fig:lrcx_d_sweep_layout}.

\begin{figure*}[htbp]
    \centering
    \includegraphics[width=\textwidth]{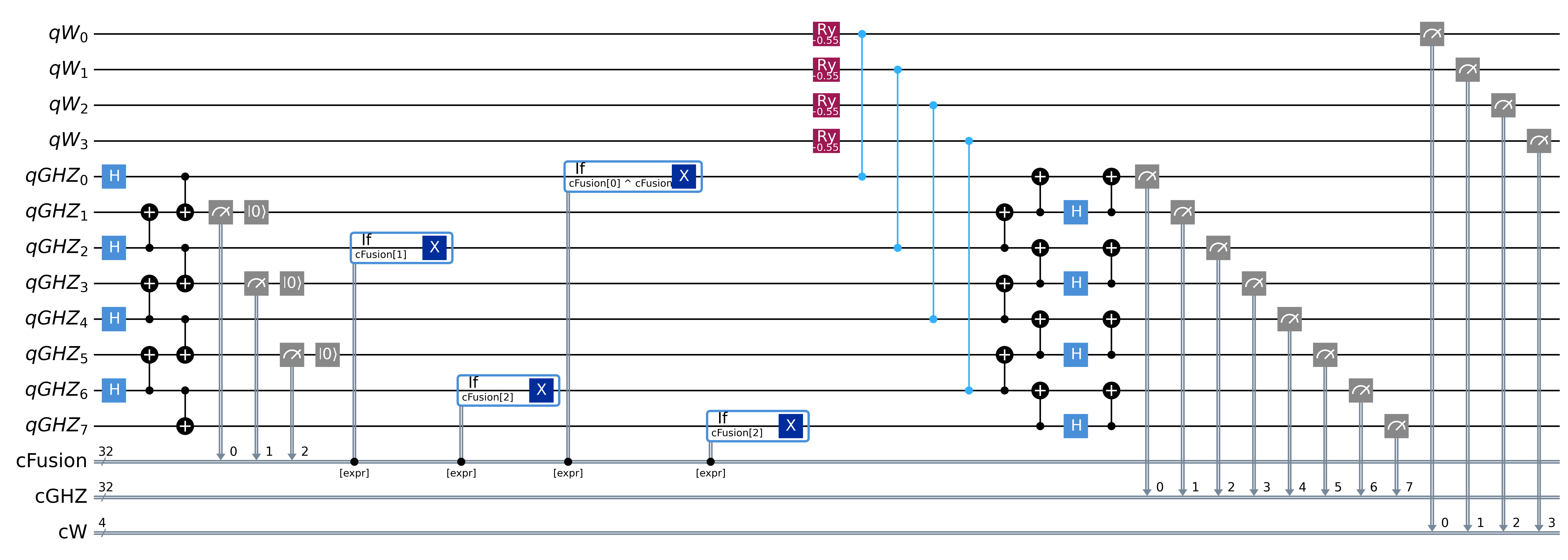}
    \caption{Error-detected, constant-depth preparation of a size $N=4$ W state using $n=2$ implicit checks. First, a GHZ state is distributed across intermediary qubits using a constant-depth dynamic circuit, then single-qubit rotations prepare a product state binomially-weighted across different excitation sectors. Finally, the GHZ state is used to carry out an error-detected, constant-depth projection onto either odd or even parity, with the former heralding successful (approximate) preparation.}
    \label{fig:wstate_circuit}
\end{figure*}

\begin{figure*}[htbp]
    \centering
    \begin{minipage}[t]{0.48\textwidth}
        \centering
        \includegraphics[width=\linewidth]{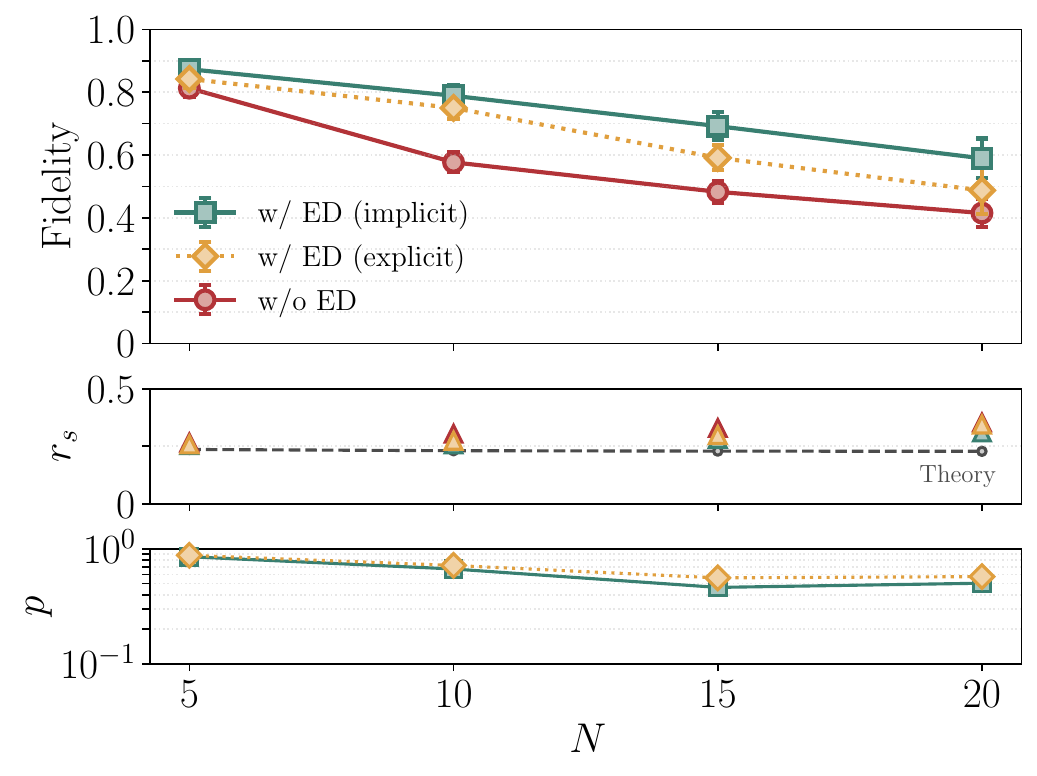}
    \end{minipage}%
    \hfill
    \begin{minipage}[t]{0.48\textwidth}
        \centering
        \includegraphics[width=\linewidth]{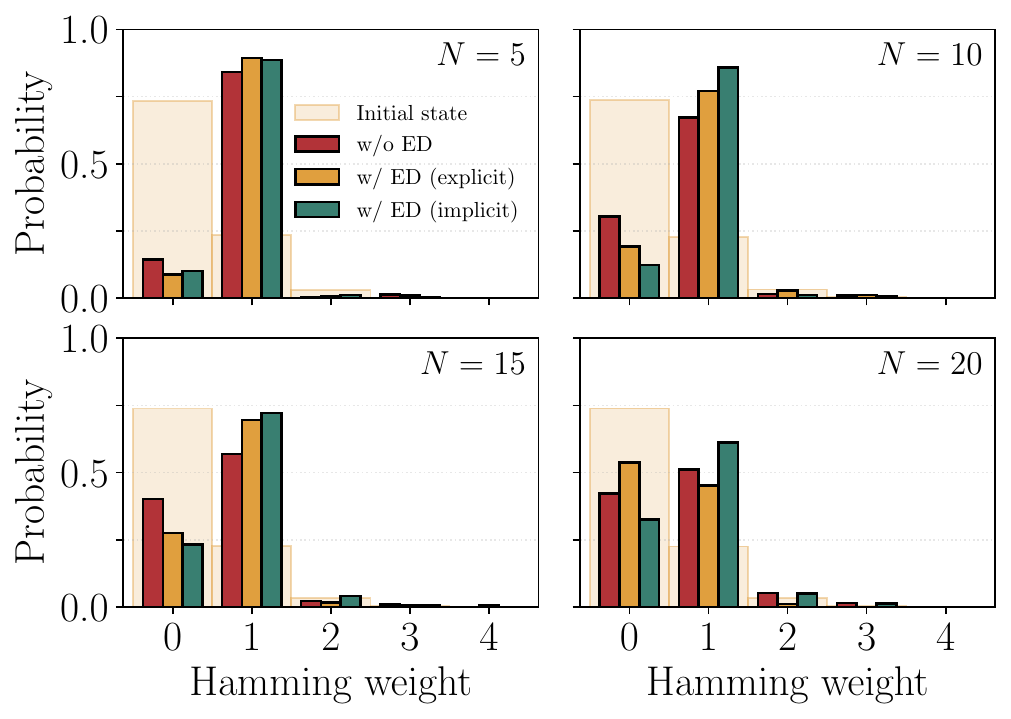}
    \end{minipage}
    \caption{W state preparation with implicit and explicit checks on \texttt{ibm\_boston}. Top left: fidelity to the ideal W state for three protocols: implicit checks ($n=2$), explicit checks ($n=2$), and no error detection ($n=1$). Middle left: Projection success rate $r_s$ for each protocol, compared with the theoretical expectation (dashed gray). Bottom left: Acceptance rate $p$ for the implicit and explicit checks. Right: Hamming weight distribution of the prepared state for each size $N$.}
    \label{fig:wstate_explicit}
\end{figure*}

For the W state experiments [Sec.~\ref{ssec:wstate}, Fig.~\ref{fig:experiment} (c) and (d)], we prepared $N$-qubit W states for $N \in \{5, 10, 15, 20\}$. An example circuit for $N=4$ with implicit checks is shown in Fig.~\ref{fig:wstate_circuit}. The qubits were laid out in a linear chain with an extra ``dangling'' qubit on every other site; preparing an $N$-qubit W state therefore requires $3N$ total physical qubits (up to 60 qubits for $N = 20$). For $N \in \{5,10,15\}$, we started with 4 different candidate layouts returned by mapomatic. We then evaluated these layouts by first computing the GHZ fidelity for the sub-circuit with the $R_y$ angles set to 0. The layouts with the best GHZ fidelity were then used to prepare the W state. For $N=20$, no layout optimization was done using GHZ fidelities. Device characterization for W state experiments at each system size are detailed in Figs.~\ref{fig:wstate_N5_layout}--\ref{fig:wstate_N20_layout}. To mitigate coherent phase errors in the intermediate GHZ state, an $R_z$ calibration scan was performed by measuring the \(\langle X X \dots X X \rangle\) for each layout prior to the main DFE measurement and $R_z$ offset angle was applied to the GHZ state preparation routine, similar to the offset used in Ref. \cite{javadi2025big}. Fidelity was estimated via direct fidelity estimation (DFE) \cite{Flammia_DirectFidelity_2011} using 2,000 shots per circuit. All common-bases observables were estimated from the same circuit run such that the  number of unique circuits for a size-$N$ W state was $N(N-1) + 1$. For instance, $N=20$ W state DFE experiment incurred 381 unique circuits and consequently $381 \times 2000 = 762,000$ shots.  Error bars were computed via bootstrap resampling with 100 iterations for 50,000 randomly sampled Pauli observables, by sampling over both the DFE measurement samples and the shot level count statistics. Fig.~\ref{fig:wstate_explicit} compares implicit and explicit checks: while both outperform the undetected baseline, implicit checks yield consistently higher fidelities across all system sizes. The explicit checks achieve a slightly higher acceptance rate $p$ than implicit checks, but this comes at the cost of reduced fidelity improvement.

\bibliography{references,extra_refs}

\end{document}